\documentclass[twocolumn]{aastex63}
\usepackage{amsmath}
\usepackage{textcomp}
\usepackage{amssymb}
\usepackage{natbib}
\usepackage{float}
\usepackage{color}
\usepackage{graphicx}
\usepackage{epstopdf}
\usepackage{mathtools}
\usepackage{xcolor}
\newcommand{\msun}{\ensuremath{M_{\odot}}}
\newcommand{\lum}{erg\,s$^{-1}$}
\newcommand{\fermi}{{\it Fermi}}
\newcommand{\nustar}{{\it NuSTAR}}
\newcommand{\nicer}{{\it NICER}}
\newcommand{\swift}{{\it Swift}}

\newcommand{\ergflux}{\mbox{${\rm \, erg \,\, cm^{-2} \, s^{-1}}$}}
\newcommand{\phflux}{\mbox{${\rm \, ph \,\, cm^{-2} \, s^{-1}}$}}
\newcommand{\gm}{$\gamma$}

\newcommand{\bzb}{BZB J0955+3551}
\newcommand{\beq}{
\begin{equation}
}
\newcommand{\eeq}{
\end{equation}
}
\newcommand{\beqa}{
\begin{eqnarray}
}
\newcommand{\eeqa}{
\end{eqnarray}
}
\newcommand{\kms}      {\ensuremath{~\mathrm{km~s^{-1}}}}
\newcommand{\msinterr} {\ensuremath{8.12 \pm 0.08}}
\newcommand{\msslopeerr} {\ensuremath{4.24 \pm 0.41}}

\submitjournal{ApJ}

\shorttitle{Candidate Neutrino Emitter BZB J0955+3551}
\shortauthors{Paliya et al.}

\begin{document}

\title{Multi-Frequency Observations of the Candidate Neutrino Emitting Blazar BZB J0955+3551}

\correspondingauthor{Vaidehi S. Paliya}
\email{vaidehi.s.paliya@gmail.com}

\author[0000-0001-7774-5308]{Vaidehi S. Paliya}
\affiliation{Deutsches Elektronen Synchrotron DESY, Platanenallee 6, 15738 Zeuthen, Germany}

\author[0000-0002-8434-5692]{M. B\"ottcher}
\affiliation{Centre for Space Research, North-West University, Potchefstroom, 2531, South Africa}

\author[0000-0001-7859-699X]{A. Olmo-Garc\'{i}a}
\affiliation{Universidad Complutense de Madrid (UCM, Spain) and Instituto de F\'{i}sica de Part\'{i}culas y del Cosmos (IPARCOS)}

\author[0000-0002-3433-4610]{A. Dom{\'{\i}}nguez}
\affiliation{IPARCOS and Department of EMFTEL, Universidad Complutense de Madrid, E-28040 Madrid, Spain}

\author[0000-0001-6150-2854]{A. Gil de Paz}
\affiliation{IPARCOS and Department of EMFTEL, Universidad Complutense de Madrid, E-28040 Madrid, Spain}

\author[0000-0002-5605-2219]{A. Franckowiak}
\affiliation{Deutsches Elektronen Synchrotron DESY, Platanenallee 6, 15738 Zeuthen, Germany}

\author[0000-0003-2403-4582]{S. Garrappa}
\affiliation{Deutsches Elektronen Synchrotron DESY, Platanenallee 6, 15738 Zeuthen, Germany}

\author[0000-0003-2434-0387]{R. Stein}
\affiliation{Deutsches Elektronen Synchrotron DESY, Platanenallee 6, 15738 Zeuthen, Germany}
\begin{abstract}
The recent spatial and temporal coincidence of the blazar TXS 0506+056 with the IceCube detected neutrino event IC-170922A has opened up a realm of multi-messenger astronomy with blazar jets as a plausible site of cosmic-ray acceleration. After TXS 0506+056, a second blazar, BZB J0955+3551, has recently been found to be spatially coincident with the IceCube detected neutrino event IC-200107A and undergoing its brightest X-ray flare measured so far. Here, we present the results of our multi-frequency campaign to study this peculiar event that includes observations with the \nustar, \swift, \nicer, and 10.4 m Gran Telescopio Canarias (GTC). The optical spectroscopic observation from GTC secured its redshift as $z=0.55703^{+0.00033}_{-0.00021}$ and the central black hole mass as 10$^{8.90\pm0.16}$~\msun. Both \nustar~and \nicer~data reveal a rapid flux variability albeit at low-significance ($\lesssim3.5\sigma$). We explore the origin of the target photon field needed for the photo-pion production using analytical calculations and considering the observed optical-to-X-ray flux level. We conclude that seed photons may originate from outside the jet, similar to that reported for TXS 0506+056, although a scenario invoking a
co-moving target photon field (e.g., electron-synchrotron)
can not be ruled out. The electromagnetic output from
the neutrino-producing photo-hadronic processes are
likely to make only a sub-dominant contribution to the
observed spectral energy distribution suggesting that the X-ray flaring event may not be directly connected with IC-200107A.

\end{abstract}

\keywords{galaxies, active --- galaxies: Evolution --- galaxies: interactions --- galaxies: jets}

\section{Introduction} \label{sec:intro}

High-energy neutrinos are unique messengers originating from the extreme physical processes in the Universe. Being solely produced in hadronic interactions of high-energy cosmic-ray nuclei with ambient matter or photon fields, they provide the smoking gun signature for hadronic acceleration sites.

Blazars, i.e., radio-loud quasars with powerful relativistic jets aligned to our line of sight, have been suggested as potential cosmic-ray and neutrino sources \citep[see, e.g.,][]{1992A&A...260L...1M,2015MNRAS.448.2412P,2017nacs.book...15M,2019ApJ...870..136L,2019ApJ...880..103G,2020ApJ...893..162F}. 
The most compelling high-energy neutrino source candidate identified so far is the blazar TXS 0506+056 \citep[][]{2018Sci...361.1378I,2018Sci...361..147I}. The 290\,TeV neutrino IC-170922A was found in spatial coincidence with TXS 0506+056 and arrived during a major outburst observable in all wavelengths \citep[][]{2018Sci...361.1378I}. Interestingly an archival search for lower-energy $\mathcal{O}$(10\,TeV) neutrinos revealed a neutrino flare in 2014/15, which lasted 160 days, but was not accompanied by activity in the electromagnetic regime \citep{2018Sci...361..147I}. From a theoretical perspective, \citet[][]{Reimer19} proposed that there should not be a strongly correlated gamma-ray and neutrino activity, and that neutrino production activity (through associated cascading) might actually show up more clearly in X-rays. However, the conclusion about the \gm-ray/PeV neutrino correlation is reported to be model-dependent \citep[see, e.g.,][]{2019ApJ...874L..29R,2020ApJ...889..118Z}.

BL Lacertae objects (or BL Lacs) are a sub-population of blazars that exhibits an optical spectrum lacking any emission lines with equivalent width $>5$\AA~\citep[e.g.,][]{1991ApJ...374..431S}. Their optical spectra are power-law dominated indicating either especially strong non-thermal continuum (due to Doppler boosting) or unusually weak thermal disk/broad line emission \citep[plausibly attributed to low accretion activity;][] {2012A&A...541A.160G}. BL Lacs that have the synchrotron peak located at very high frequencies ($\nu^{\rm peak}_{\rm syn}\geq10^{17}$ Hz) are termed as extreme blazars \citep[e.g.,][]{2001A&A...371..512C,2019MNRAS.486.1741F,2019ApJ...882L...3P}. The observation of such a high synchrotron peak frequency indicates them to host some of the most efficient particle accelerator jets. Interestingly, extreme blazars are also proposed as promising candidates of high-energy neutrinos \citep[cf.][]{2015MNRAS.448.2412P,2016MNRAS.457.3582P}.

So far, any clustering of neutrinos in either space or time have not been confirmed in the all-sky searches of IceCube data \citep[][]{2015ApJ...807...46A,2017ApJ...835..151A,Aartsen:2019fau}. Therefore, a promising methodology could be the search for transient and variable electromagnetic sources temporally and spatially coincident with IceCube neutrino events using multi-frequency observations. 

In this regard, the identification of a \gm-ray detected extreme blazar, \bzb~(also known as 4FGL J0955.1+3551), found in 
spatial coincidence with the IceCube detected neutrino event IC-200107A \citep[][]{2020GCN.26655....1I,2020ATel13394....1G,2020ATel13395....1K} has provided an interesting case for blazar jets as a plausible source of cosmic neutrinos. In fact, a prompt \swift-XRT target of opportunity (ToO) observation of \bzb~on 2020 January 8 found it to be undergoing its brightest X-ray flare measured so far. Another \gm-ray detected blazar, 4FGL J0957.8+3423, was found to lie within the 90\% positional uncertainty of IC-200107A, however, no significant flux enhancement was noticed from this object in X- or \gm-rays \citep[][]{2020ATel13395....1K,2020GCN.26669....1G}.

Motivated by the identification of a candidate neutrino emitting blazar undergoing an X-ray outburst close in time to the neutrino arrival, we started a multi-wavelength campaign. This includes a Director's Discretionary Time (DDT) observation with the Nuclear Spectroscopic Telescope Array (\nustar) and multiple \swift~target of opportunity (ToO) observations. An optical spectroscopic followup with 10.4 m Gran Telescopio Canarias (GTC) was carried out to determine the spectroscopic redshift of \bzb. In addition to that, the source was also observed with the Neutron star Interior Composition Explorer (\nicer) simultaneous to the \nustar~pointing as a part of DDT ToO invoked by the mission principal investigator. Here, we present the results of the conducted multi-frequency campaign and attempt a theoretical interpretation to understand the underlying physical processes. In Section~\ref{sec:analysis}, we describe the steps adopted to analyze various data sets. Results are presented in Section~\ref{sec:results} and discussed in Section~\ref{sec:discussion}. We summarize our findings in Section~\ref{sec:summary}. Throughout, we adopt a Cosmology of $H_0=67.8$~km~s$^{-1}$~Mpc$^{-1}$, $\Omega_m = 0.308$, and $\Omega_\Lambda = 0.692$ \citep[][]{2016A&A...594A..13P}.

\section{Data Reduction and Analysis}\label{sec:analysis}
\subsection{Optical Spectroscopy with GTC}
The $i'$ filter image of \bzb~taken with the Panoramic Survey Telescope and Rapid Response System (Pan-STARRS) is shown in Figure~\ref{fig:optical}. A faint companion object ($i'$ magnitude = 20.85$\pm$0.36) located $\sim$3$^{\prime\prime}$ South-East of the blazar ($i'$ magnitude = 19.17$\pm$0.06) can be seen. Since both objects lacked spectroscopic redshift information, we carried out a long-slit spectroscopy of the system with Optical System for Imaging and low-Intermediate-Resolution Integrated Spectroscopy \citep[OSIRIS;][]{2000SPIE.4008..623C,2003SPIE.4841.1739C} spectrograph mounted at GTC.

The 0.8 arcsec-wide slit was positioned to cover the source and a companion $\sim$3 arcsec south-east to the blazar (see Figure~\ref{fig:optical}). The total integration time was $\sim$2~hrs divided into 6 exposures of 1098 s each. The chosen grism was R1000R, which covers the spectral range of  5100 - 10000~\AA\ with a resolution ($\lambda/\Delta\lambda$) of 1122\footnote{\url{http://www.gtc.iac.es/instruments/osiris/osiris.php}}. This grism was selected due to its large spectral range and good spectral resolution, which provides a large pool to find  emission or absorption lines and calculate the redshift of the source.

The raw data was reduced using the standard procedure with the IRAF tasks, through the PyRAF software\footnote{PyRAF is a product of the Space Telescope Science Institute, which is operated by AURA for NASA \url{ http://www.stsci.edu/institute/software_hardware/pyraf/}}. The main steps are: bias and flat correction, cosmic-ray removal, wavelength calibration, sky subtraction, spectra extraction and flux calibration. 
The cosmic rays were removed in each individual science spectrum with the IRAF task {\tt lacos\_spec} \citep{2001PASP..113.1420V}. The wavelength calibration was done with a combination of arcs from three different lamps (Hg-Ar, Ne and Xe) to cover all the wavelength range of the spectra. The sky was subtracted with the IRAF task {\tt background}, selecting background samples to the right of the blazar and to the left on the companion, and fitted with a Chebyshev polynomial of order 3. After this step, the science spectra were combined, which removed any cosmic-ray residual.  The spectrum of the blazar and the companion were extracted independently from the combined science spectra. The extraction was done with the IRAF task {\tt apall}, optimising the apertures to extract the most flux from the sources. For the flux calibration, the spectrophotometric standard star G191-B2B was observed on the same night of the observation. This  calibration included atmospheric extinction correction  at the observatory \citep{extinction}.  Each spectrum was flux calibrated to convert from counts to absolute flux units, and corrected from Galactic extinction using the IRAF task {\tt deredden} with the values $\mathrm{R} = 3.1$, $\mathrm{E(B-V)}= 0.0109$ \citep{2011ApJ...737..103S}. 

\subsection{\nustar, NICER, and \swift}
\nustar~observed \bzb~on 2020 January 11 for a net exposure of 25.6 ksec under our DDT request (observation id: 90501658002, PI: Paliya). We first cleaned and calibrated the event file using the tool {\tt nupipeline}. We define the source and background regions as circles of 30$^{\prime\prime}$ and 70$^{\prime\prime}$, respectively. The former was centered at the target blazar and the latter from a nearby region on the same chip and avoiding source contamination. The pipeline {\tt nuproducts} was used to extract light curves, spectra, and ancillary response files. In the energy range of 3$-$79 keV, a binning of 1.5 ksec was adopted to generate the light curve and the source spectrum was binned to have at least 20 counts per bin.

\nicer~observed \bzb~ for a net exposure of $\sim$11 ksec, simultaneous to the \nustar~pointing on 2020 January 11 as a DDT target of opportunity (ToO observation id: 2200990102). We analyzed the \nicer~data with the latest software HEASOFT 6.26.1 and calibration files (v. 20190516). In particular, the pipeline {\tt nicerl2} was adopted with default settings to selects all 56 detectors, apply standard filters and calibration to clean the events and finally merge them to generate one event file. We then used the tool {\tt xselect} to extract the source spectrum and 3 minutes binned light curve. The background was estimated using the tool {\tt nicer\_bkg\_estimator}\footnote{\url{https://heasarc.gsfc.nasa.gov/docs/nicer/tools/nicer\_bkg\_est\_tools.html}} (K. Gendreau et al. in preparation). The quasar spectrum was binned to 20 counts per bin. 

Close in time to the arrival of IC-200107A neutrino, ToO observations of \bzb~from the \swift~satellite were carried out on 2020 January 8 \citep[][]{2020ATel13394....1G,2020ATel13395....1K}, 10 and 11. We first cleaned and calibrated the X-ray Telescope (XRT) data taken in the photon counting mode with the tool {\tt xrtpipeline} and by adopting the latest CALDB (v. 20200106). Exposure maps and ancillary response files were generated with the tasks {\tt ximage} and {\tt xrtmkarf}, respectively. To extract the source spectrum, we considered a circular region of 47$^{\prime\prime}$, which encloses about 90\% of the XRT point spread function, centered at the target. The background was estimated from an annular region centered at the target with inner and outer radii 70$^{\prime\prime}$ and 150$^{\prime\prime}$, respectively. We binned the blazar spectrum to 20 counts per bin. The X-ray spectral analysis was carried out in XSPEC \citep[][]{Arnaud96} and the Galactic neutral hydrogen column density ($N_{\rm H}=1.14\times10^{20}$ cm$^{-2}$) was adopted from \citet[][]{Kalberla05}.

Individual snapshots taken from the \swift~UltraViolet Optical Telescope (UVOT) were first combined using the pipeline {\tt uvotimsum} and then photometry was performed with the task {\tt uvotsource}. For the latter, we considered a source region of 2$^{\prime\prime}$, avoiding the nearby object located $\sim$3$^{\prime\prime}$ South-East of \bzb. The background is estimated from a 30 $^{\prime\prime}$ circular region free from the source contamination. The derived magnitudes were corrected for Galactic extinction \citep[][]{2011ApJ...737..103S} and converted to flux units following zero points adopted from \citet[][]{2011AIPC.1358..373B}.

\subsection{Others}
\bzb~remained below the detection threshold of the \fermi-Large Area Telescope at the time of the neutrino arrival and prior on month-to-years timescale \citep[][]{2020GCN.26669....1G}. Therefore, we used the spectral parameters provided in the recently released fourth catalog of the \fermi-LAT detected objects \citep[4FGL;][]{2020ApJS..247...33A} to get an idea about the average \gm-ray behavior of the source. In addition to that, we used archival measurements from the Space Science Data Center\footnote{\url{https://tools.ssdc.asi.it/}}. These datasets can provide a meaningful information about the typical activity state of the source.

\subsection{Probability of chance coincidence}
The third catalog of high-synchrotron peaked blazars \citep[3HSP;][]{2019A&A...632A..77C} contains 384 extreme blazars, yielding a density of $9.3 \times 10^{-3}$ per sq. deg. of sky. Since the total number of extreme blazars are predicted to be $\sim$400 \citep[][]{2019A&A...632A..77C}, the sample of extreme blazars present in the 3HSP catalog can be considered almost complete. Given that IC200107A had a 90\% localization of 7.6 deg, we thus expect to find $7.1 \times 10^{-2}$ extreme blazars coincident with the neutrino. 

We can additionally determine the X-ray flare rate or duty cycle (DC) for extreme blazars, as for any other class of astrophysical objects, using the X-ray variability information collected from an all-sky surveying instrument. For this we used publicly available 2$-$12 keV light curves generated using the data from the All Sky Monitor (ASM) on board the \textit{Rossi} X-ray Timing Explorer (RXTE) mission\footnote{\url{http://xte.mit.edu/ASM\_lc.html}}. We cross-matched the RXTE ASM catalog of 587 sources with 3HSP, and found ASM light curves for 17 extreme blazars. To avoid spurious detection due to poor sensitivity of the instrument, for each object, we considered only data points which qualified the following two filters: (i) the count rate ($R_{\rm ASM}$) should be positive, and (ii) $R_{\rm ASM}/\Delta R_{\rm ASM}>$2, where $\Delta R_{\rm ASM}$ is the 1$\sigma$ uncertainty in $R_{\rm ASM}$. Furthermore, the observation on a particular day was considered as a flare if the count rate estimated for that day of observation ($R_{\rm ASM,i}$) fulfilled the following condition:

\begin{equation}
    R_{\rm ASM,i} - \Delta R_{\rm ASM,i} \geq 2\times \langle R_{\rm ASM} \rangle
\end{equation}\label{eq:dc}

\noindent where $\langle R_{\rm ASM} \rangle$ is the median count rate for the mission light curve. If the observation on a particular day qualified the above mentioned filter (Equation~\ref{eq:dc}), we flagged it as a `flare', otherwise `non-flare'. The DC is then computed as the ratio of the number of flaring epochs divided by total observing epochs. This exercise led to the mean DC for the sample as 5.2\% with a range of 1.7-8.9\%. Assuming the mean DC of these 17 extreme blazars is representative of the broader extreme blazar population, the probability of finding a coincident extreme blazar by chance that is simultaneously flaring in X-rays is just $3.7 \times 10^{-3}$. This estimate is, however, specific to IC200107A. The possibility that other high-energy neutrinos may have had flaring extreme blazar counterparts is difficult to quantify without a systematic follow-up program.

\subsection{Neutrino flux estimate}\label{sec:nu_fl}
A single high-energy neutrino detection from the extreme blazar population would suggest a cumulative expectation of 0.05 $< N_{\textup{pop}} < $ 4.74 at 90\% confidence, with each of the 384 extreme blazars contributing some fraction of this total \citep{Strotjohann19}. If each had an equal likelihood to generate a neutrino alert, then we would expect $ 1.3 \times 10^{-4} \lesssim N_{\textup{src}} \lesssim 0.012$ per extreme blazar.

Given that the association with \bzb~ is not dependent on the event topology of IC200107A, we simply require sufficient neutrino flux for a single high-energy neutrino alert under any of the public IceCube realtime alert selections. IC200107A was identified by a new neural network classifier \citep{2020GCN.26655....1I}, which identifies high-energy starting track events with high efficiency \citep{2019ICRC...36..937K}. However, with an overall rate of high-energy starting tracks that is just $\sim$2 per year \citep{2020GCN.26655....1I}, the effective area for this selection is still substantially smaller than that for through-going muon alerts \citep{2019ICRC...36.1021B}. 

We can derive the necessary neutrino fluence normalization taking the sum of neutrino effective areas at the declination of \bzb~ over the duration of the Icecube Realtime System. For this 4 year period, which overlaps a transition in IceCube event selections, we integrate each effective area over the period that they were active  \citep{2017APh....92...30A, 2019ICRC...36.1021B}. No neutrino energy estimate was provided for IC200107A \citep{2020GCN.26655....1I}, so we here assume an approximate neutrino energy of $\sim$100 TeV, the energy at which most starting tracks are expected for an $E^{-2}$ spectrum. The effective area at this energy was  0.7 m$^{2}$ under the old alert selection\citep{2017APh....92...30A}, and 9.48 m$^{2}$ under the new alert selection \citep{2019ICRC...36.1021B}, yielding a weighted average of 2.9 m$^{2}$ at the declination of the source. With this effective area, at 100 TeV, we require a mean neutrino flux of $6 \times 10^{-15} < F_{\textup{steady}} < 5 \times 10^{-13}$ \ergflux~for extreme blazars such as \bzb. If we assume neutrino emission from these sources is dominated by X-ray flares, then for a DC of 5.2\% we expect a flux of $1 \times 10^{-13} < F_{\textup{flare}} < 1 \times 10^{-11}$ erg cm$^{-2}$ s$^{-1}$ for the duration of each flare.
Given the large range of the expected neutrino flux, we conservatively assume a value of 10$^{-13}$ \ergflux~for rest of the calculation, considering an Eddington bias of a factor of 100 \citep[][]{Strotjohann19}.

\section{Results}\label{sec:results}
\begin{figure*}[t!]
\centering
\hbox{
\includegraphics[scale=0.37]{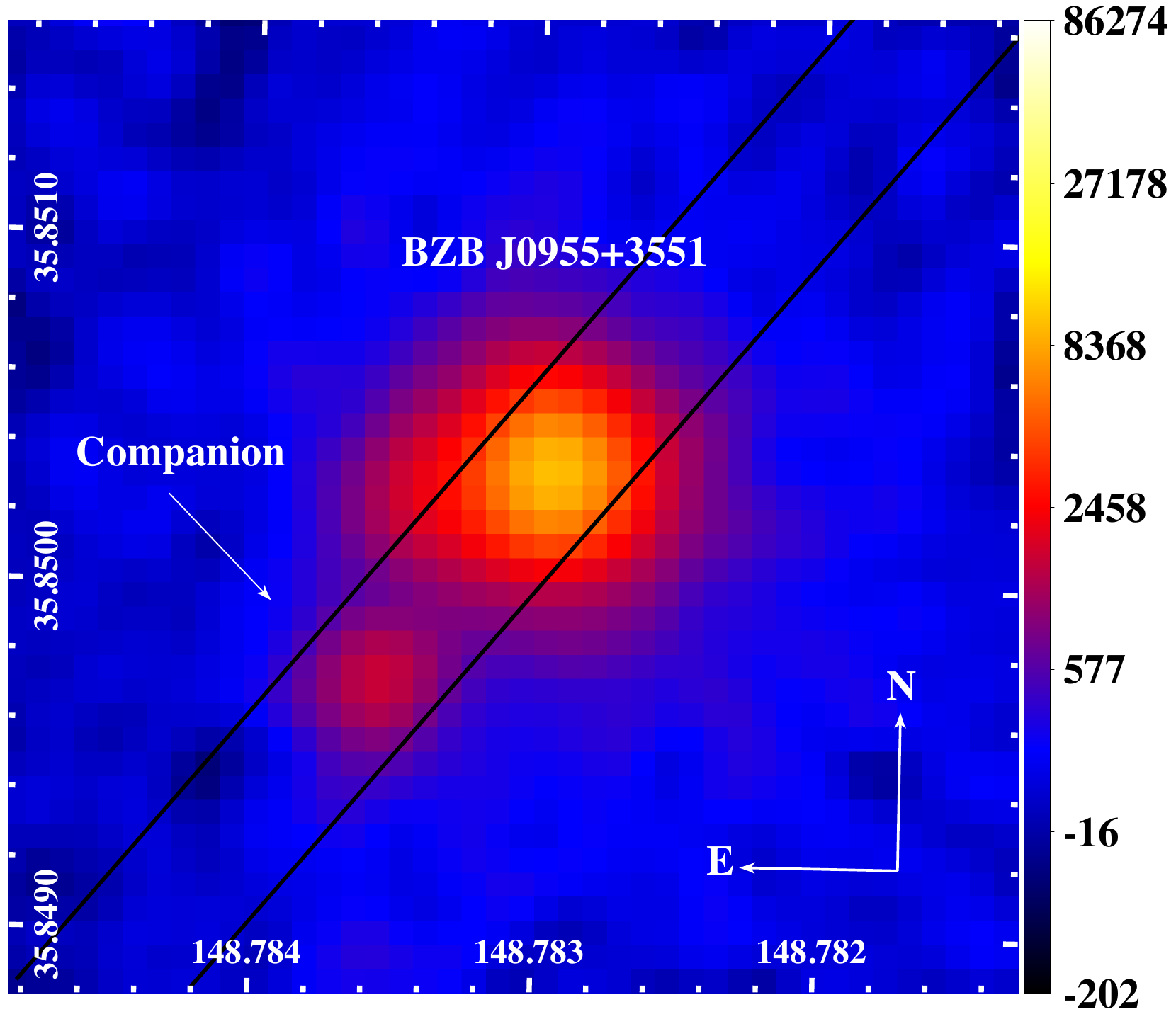}
\includegraphics[scale=0.37]{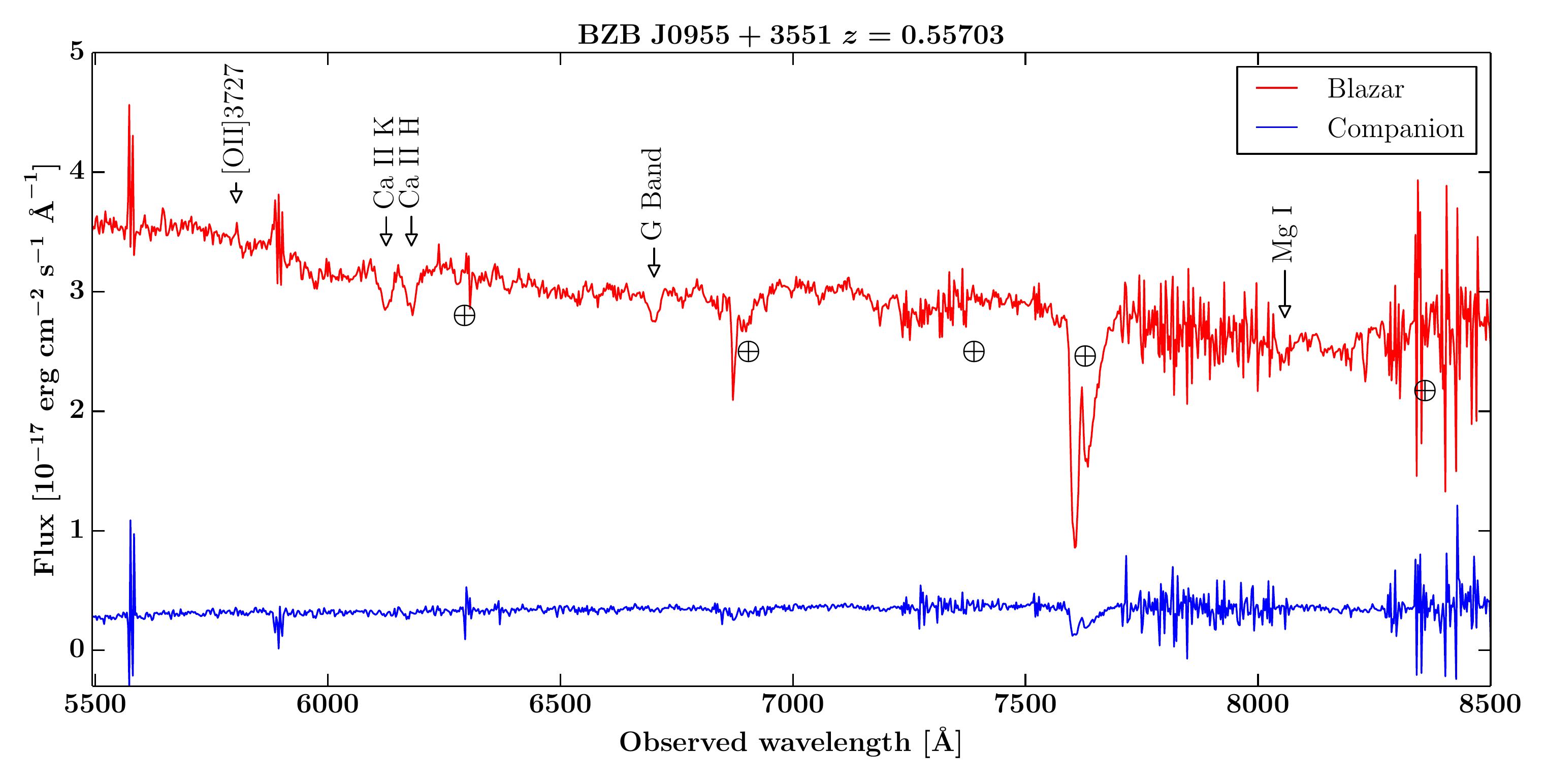}
}
\caption{Left: Pan-STARRS $i'$ filter image of \bzb. Note the presence of a faint companion object at $\sim$3$^{\prime\prime}$ South-East of the blazar. Parallel black lines represent the slit position for the long-slit spectrograph OSIRIS. The colorbar represents the Pan-STARRS count units. Right: Optical spectra of the source \bzb\ and the companion taken at GTC with OSIRIS. The red line is the spectrum of the blazar and the blue line is the spectrum of the companion. The identified emission and absorption lines are marked with vertical arrows and labeled correspondingly. The atmospheric absorption features are marked with the symbol $\oplus$. 
}
\label{fig:optical}
\end{figure*}

\begin{figure*}[t!]
\hbox{\hspace{0.0cm}
\includegraphics[scale=0.43]{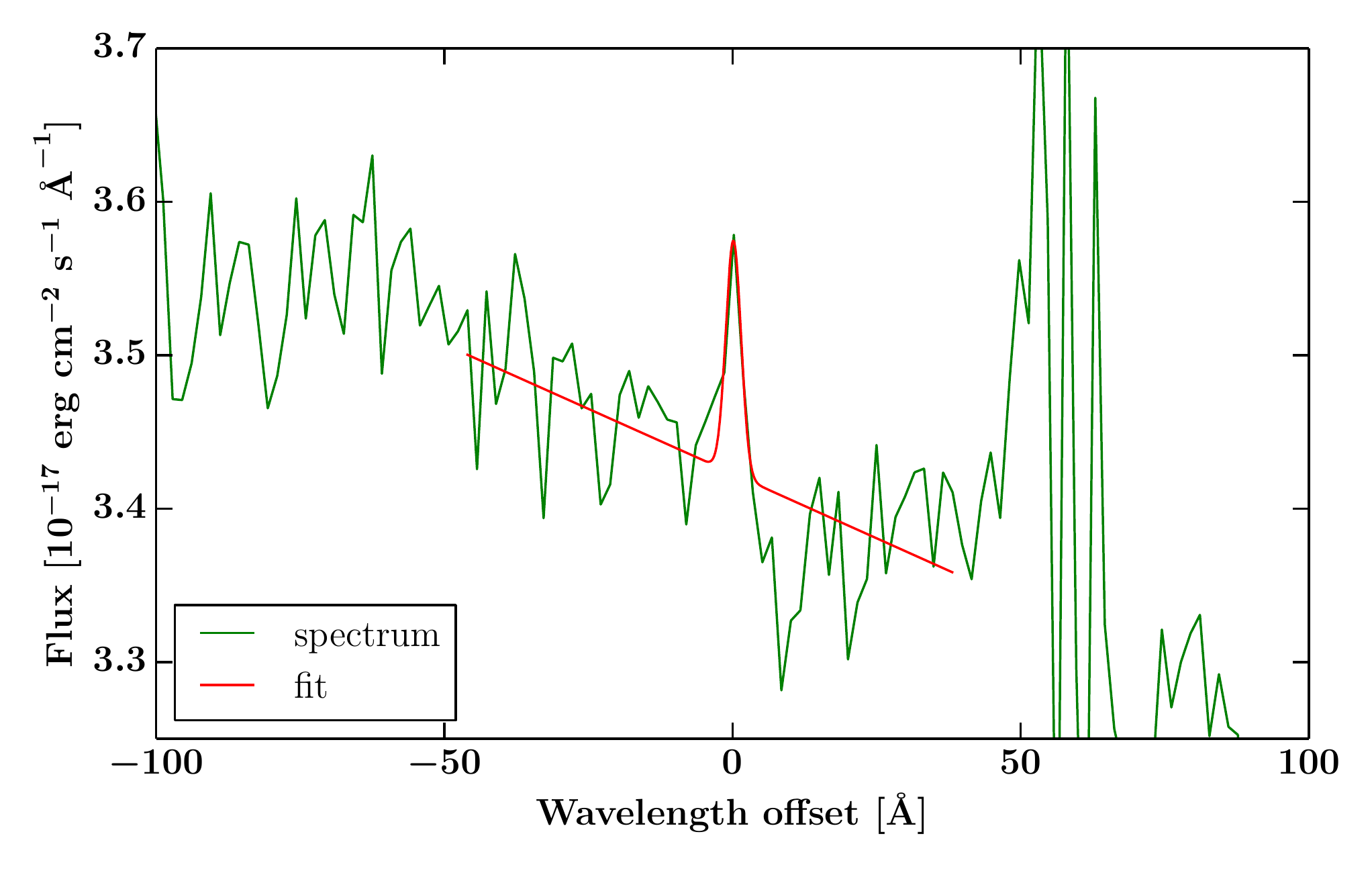} 
\includegraphics[scale=0.46]{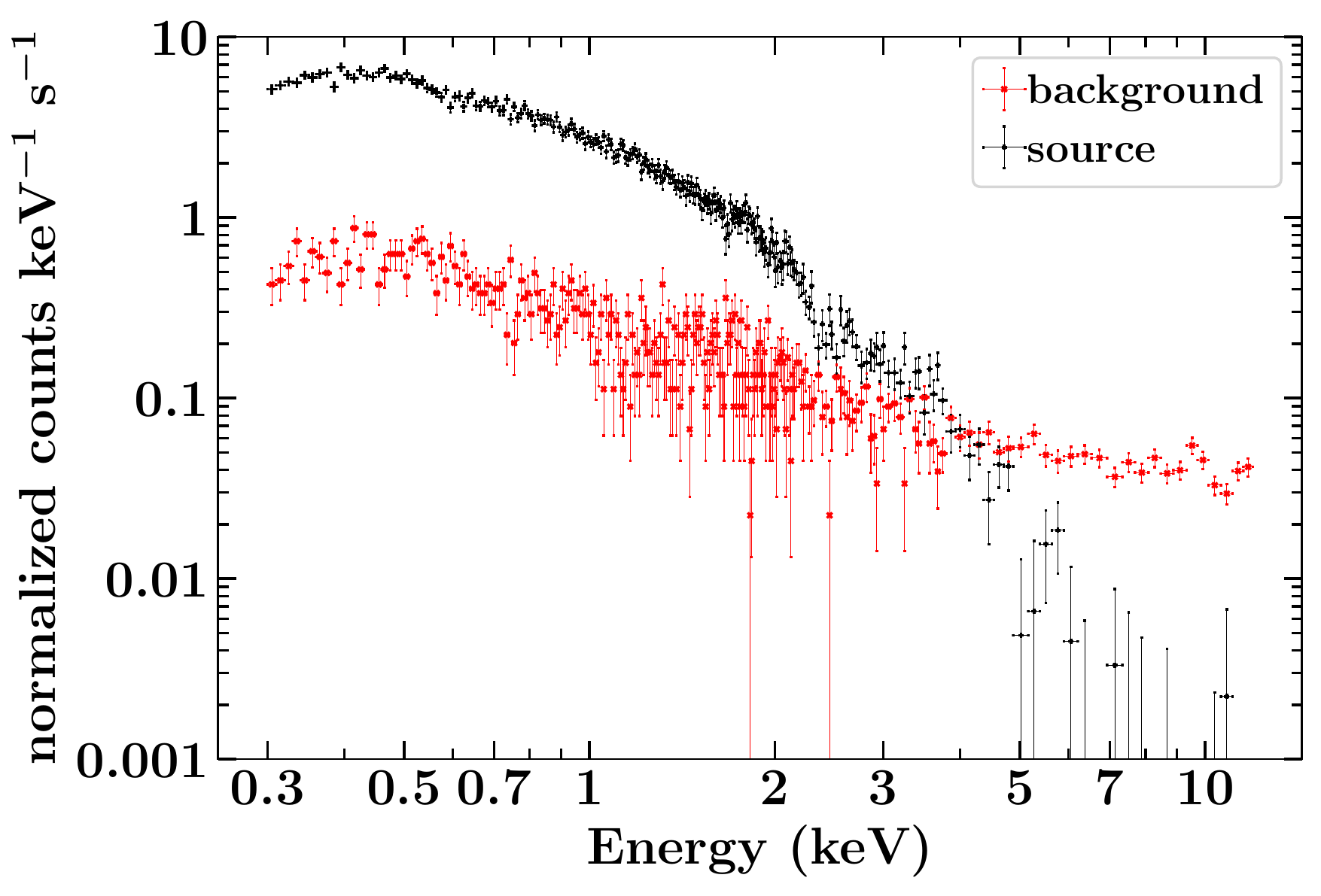} 
}
\hbox{
\includegraphics[scale=0.5]{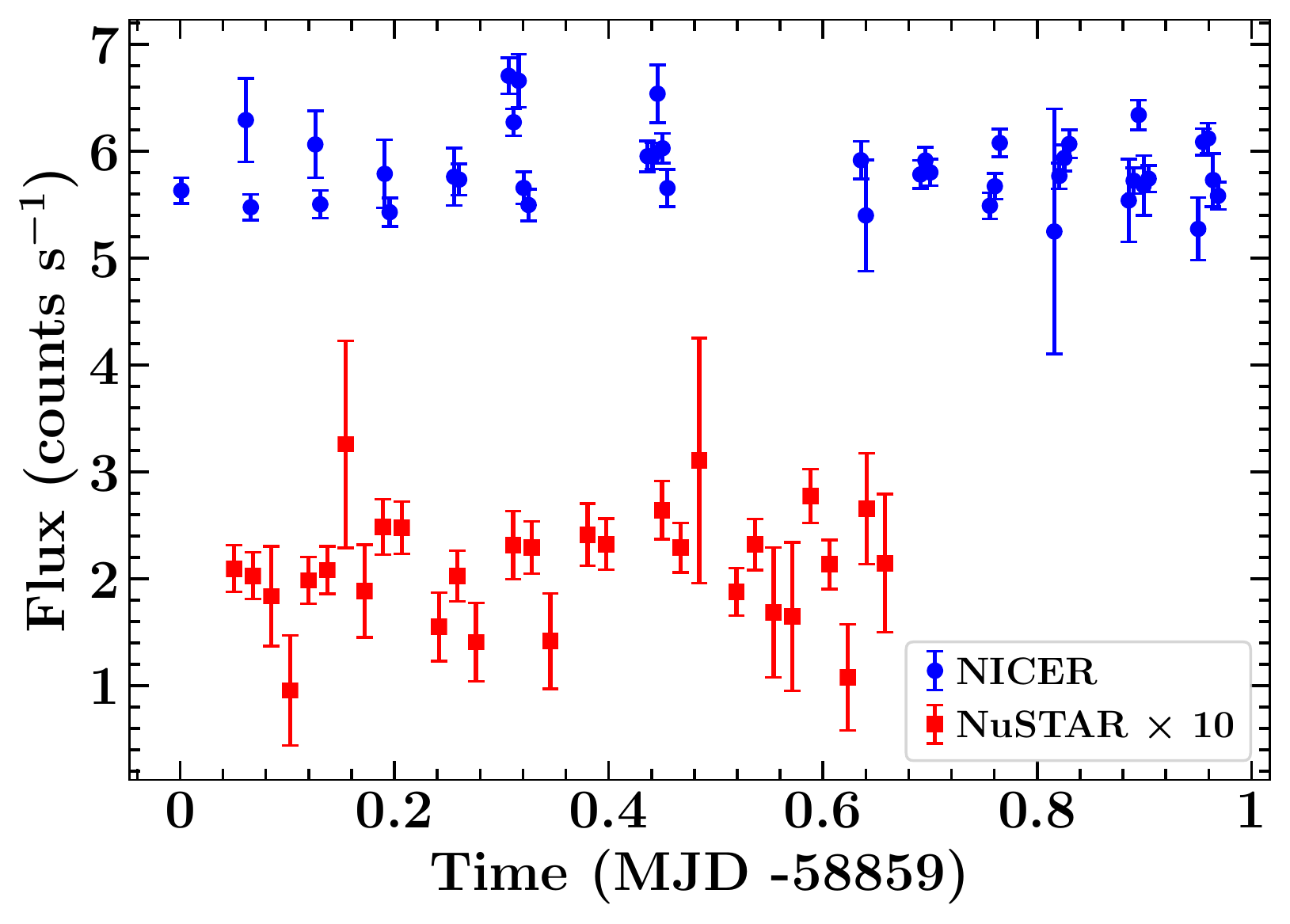} 
\includegraphics[scale=0.5]{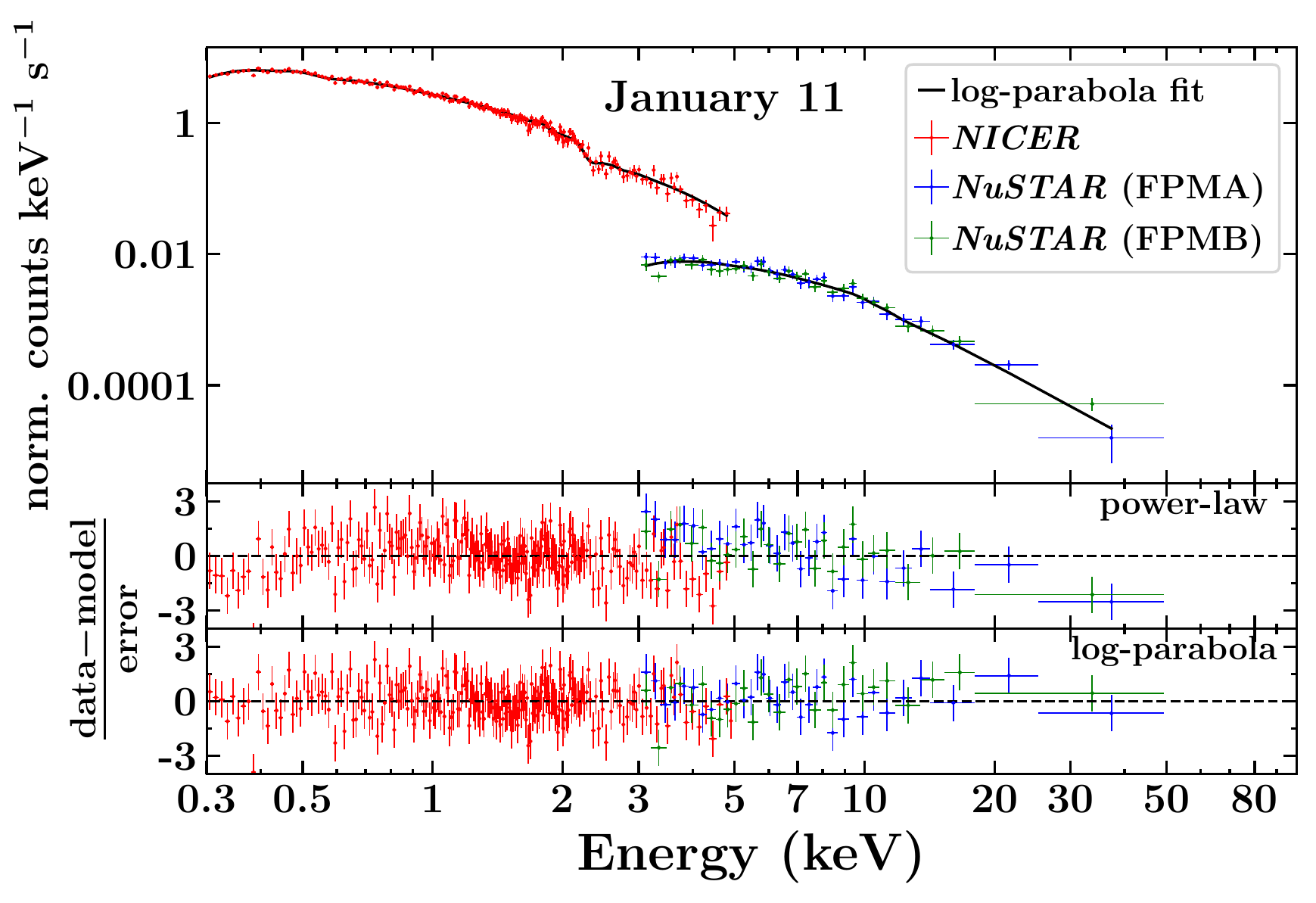} 
}
\caption{Top left: Fit of the emission line [OII]3727 to derive its luminosity and equivalent width. The spectrum is shown in a green line and the fit in a red line. The continuum is fit with a polynomial of degree 1 and the line with a single Gaussian function. The wavelength of the line has been subtracted from the x-axis. Top right: \nicer~count spectra of \bzb~(black) and the background (red). As can be seen, at $\gtrsim5$ keV, the background dominates the observed counts. Bottom left: 1.5 ksec binned \nustar~(3$-$79 keV) and 3 minutes binned \nicer~(0.3$-$5 keV) light curves of \bzb. The flux of the former is multiplied by 10 for a meaningful comparison. Bottom right: The \nicer~and \nustar~counts spectrum jointly fitted with the best-fitted log-parabola model. Lower panels show the residuals of the fit for two models: power-law and log-parabola, as labeled.}
\label{fig:x-ray}
\end{figure*}

The optical spectra of \bzb~and the nearby companion are shown in Figure~\ref{fig:optical}. Various absorption lines associated with the host galaxy, e.g., Ca II H\&K doublet, are identified in the optical spectrum of the blazar. Additionally, we also detected a weak [O II]3727 emission line. These allowed us to firmly establish the redshift of \bzb~as $z=0.55703^{+0.00033}_{-0.00021}$. The spectrum of the companion does not reveal any noticeable feature. Deeper spectroscopic observations are necessary to characterize this object and to explore the possibility of its interaction/merger with \bzb.

We computed the rest-frame equivalent width of the [OII]3727 emission line by fitting the continuum around the emission line with a polynomial of degree 1 and the emission line with a Gaussian function (see top left panel in Figure~\ref{fig:x-ray}). The data was normalized by the factor $10^{-17}$. We first fitted the continuum with a sample of 45 points, $\sim$30 and $\sim$40 \AA\, to either side of the line. With the continuum subtracted from the spectrum, we fitted the emission line using 7 points ($\sim 11$\AA), more than the three free parameters in the fit.  This leads to the rest-frame equivalent width of 0.15$\pm$0.05 \AA~and line luminosity as $(6\pm 2)\times10^{39}$ \lum. Note that the signal-to-noise ratio around [OII]3727 line is $>$70 which ensures that the estimated values are reliable. Moreover, during the analysis, we varied the extraction aperture, which changed the amount of sky residuals in the final spectrum, to determine if the observed emission line could be due to background noise. In all cases, the line was clearly visible. Therefore, we conclude that the line detection is real and free from any artifacts.

In order to ascertain the impact of the background on the \nicer~observation, we plot the count spectrum of the source and background in Figure~\ref{fig:x-ray} (top right panel). As can be seen, the \nicer~spectrum remains source dominated up to $\sim$5 keV. Therefore, we used 0.3$-$5 keV energy range to extract the \nicer~light curve and spectrum of \bzb.

In the bottom left panel of Figure~\ref{fig:x-ray}, we show the \nicer~and the \nustar~light curves. The light curves are scanned to search for rapid flux variations. This was done by computing the flux doubling/halving time ($\tau$) as follows:

\begin{equation}\label{eq:flux_double}
F(t_2) = F(t_1) 2^{(t_2-t_1)/\tau}
\end{equation}

\noindent where $F(t_1)$ and $F(t_2)$ are the fluxes at time $t_1$ 
and $t_2$ respectively. The uncertainties in the flux values were taken into account by setting the conditions that the difference in fluxes at the epochs $t_1$ and $t_2$ is at least 2$\sigma$ significant.

\begin{figure*}[t!]
\hbox{\hspace{0.0cm}
\includegraphics[scale=0.57]{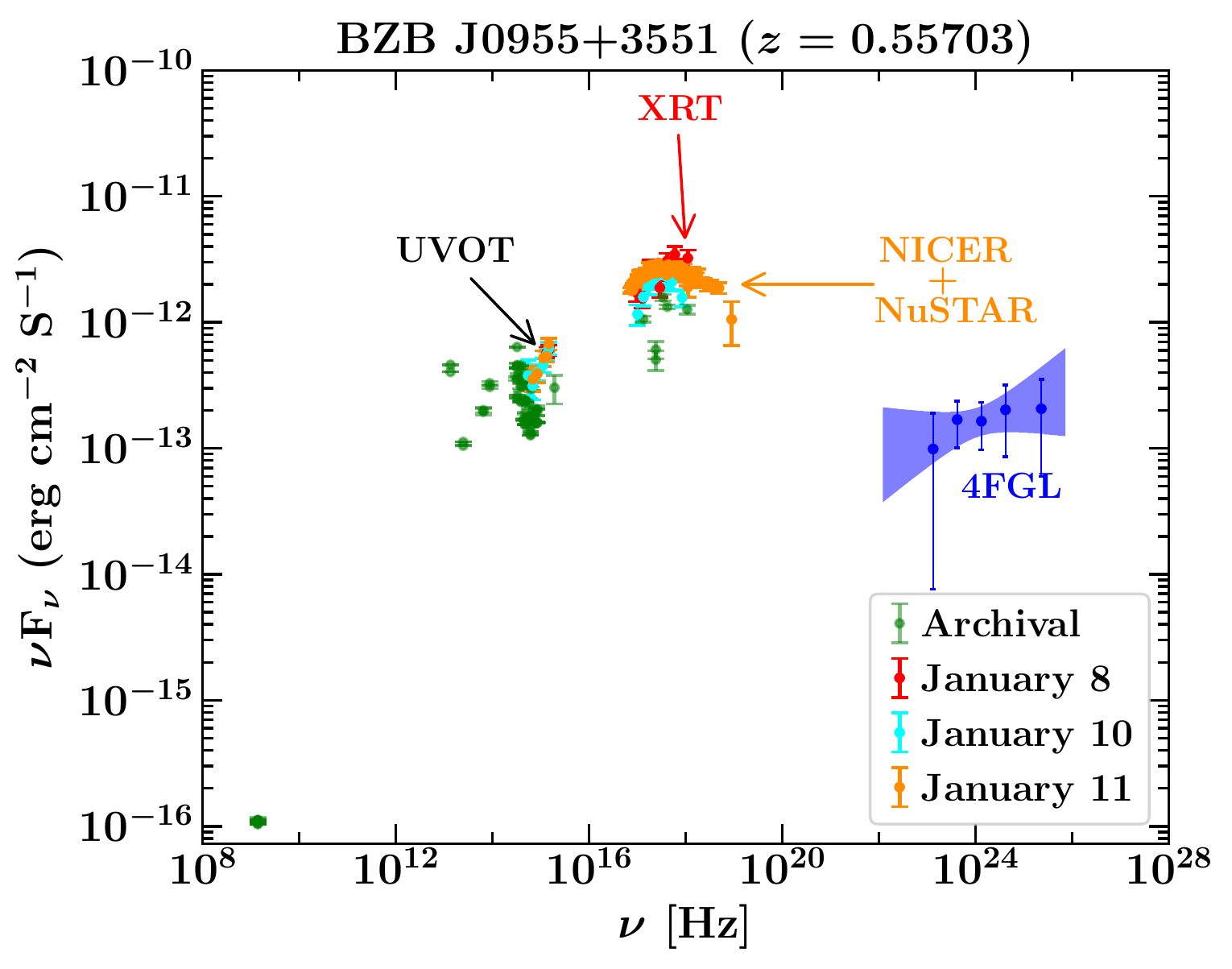} 
\includegraphics[scale=0.57]{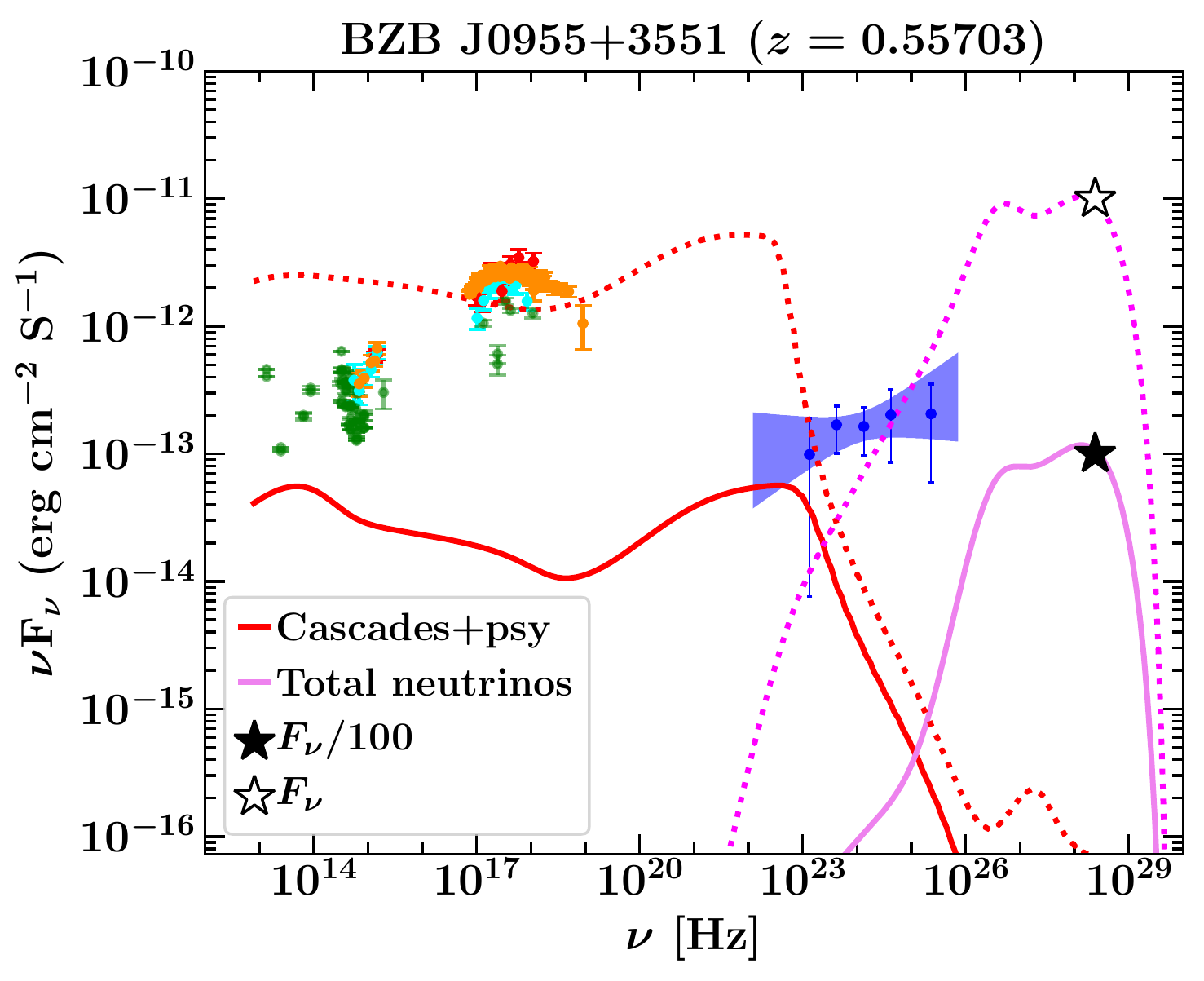} 
}
\caption{Left: Broadband SED of \bzb~generated using the data acquired on January 8 (red) and 11 (yellow) and also considering archival observations (green). In the \fermi-LAT energy range, we show the bow-tie and spectral data points adopted from the 4FGL catalog. Right: Same as left but plotting the results of the hadronic simulation performed using the parameters constrained from the observed optical-to-X-ray spectrum and derived from the analytical calculations in Section~\ref{sec:general}. We divide the expected 100 TeV neutrino flux of $1\times10^{-11}$ \ergflux~(black empty star) by a factor of 100 (black filled star) to take into account the Eddington bias \citep[cf.][]{Strotjohann19}.}
\label{fig:sed}
\end{figure*}
\begin{table*}
\begin{center}
\caption{Summary of the SED analysis.\label{tab:sed_par}}
\begin{tabular}{lccccccc}
\tableline\tableline
&  & & & X-ray & & &\\
 Epoch & $\Gamma_{\rm X}/\alpha$ & $\beta$ & Normalization & Flux & $chi^2$/dof & Prob. & Instrument\\
\tableline
 January 8 & 1.74$^{+0.11}_{-0.10}$ & -- & 14.11$^{+1.03}_{-1.03}$ & 5.37$^{+0.97}_{-0.77}$ & 22.93/27 & 0.6 & \swift-XRT\\
 January 10 & 2.83$^{+0.44}_{-0.38}$ & 1.10$^{0.54}_{0.49}$ & 1.17$^{+0.21}_{-0.22}$ & 1.89$^{+0.67}_{-0.48}$ & 14.07/21 & $<10^{-4}$ & \swift-XRT\\
 January 11 & 2.13$^{+0.04}_{-0.04}$ & 0.17$^{+0.04}_{-0.03}$ & 1.74$^{+0.08}_{-0.07}$ & 3.80$^{+0.14}_{-0.17}$ & 456.94/462 & $<10^{-4}$ &\nicer+\nustar\\
            &                        &                        &                        & 2.11$^{+0.21}_{-0.15}$ &            &  \\
 \tableline
 & & & \swift-UVOT & & & & \\
 Epoch & $V$ & $B$ & $U$ & $UVW1$ & $UVM2$ & $UVW2$\\
 \tableline
 January 8 & -- & -- & -- & -- & 5.75$\pm$0.56 & 6.03$\pm$0.54 & \\
 January 10 & 3.78$\pm$1.24 & 3.13$\pm$0.70 & 4.03$\pm$0.59 & 4.58$\pm$0.58 & 5.54$\pm$0.55 & 6.23$\pm$0.70 & \\
 January 11 & -- & 3.54$\pm$0.72 & 3.91$\pm$0.58 & 5.20$\pm$0.72 & 5.36$\pm$0.51 & 6.75$\pm$0.74 & \\
 \tableline
\end{tabular}
\end{center}
\tablecomments{\swift-XRT spectral fitting (January 8 and 10) is done in the energy range of 0.3$-$10 keV, whereas, it is 0.3$-$79 keV for the joint \nicer~and \nustar~analysis (January 11). $\Gamma_{\rm X}$ is the power-law X-ray photon index and $\alpha$ and $\beta$ are the log-parabolic photon index at the pivot energy (fixed at 3 keV) and curvature around the peak, respectively. The X-ray normalization has the unit of 10$^{-4}$ \phflux~keV$^{-1}$. The quoted flux values are in 2$-$10 keV energy range and in the second row of January 11 data, we also provide the flux in 10$-$79 keV. The power-law and log-parabola models were compared by adopting f-test and the derived probability of null-hypothesis (that the power-law model is a better representation of the data) is given in the column `Prob.'. The flux values reported for the \swift-UVOT filters are in 10$^{-13}$ \ergflux~and are corrected for Galactic reddening.}
\end{table*}

We found evidence of rapid flux variations in the \nicer~data with the shortest flux halving time of 28.3$\pm$7.9 minutes at 3.5$\sigma$ significance level. The \nustar~light curve also revealed traces of fast variability with the shortest flux doubling time of 19.2$\pm$10.7 minutes, albeit at a low 2.2$\sigma$ confidence level.

In order to search for curvature in the X-ray spectrum, we fitted two models, a power-law and a log-parabola, taking into account the Galactic absorption. The goodness of the fit was determined using f-test. The results  of the spectral analysis are provided in Table~\ref{tab:sed_par} and residuals of the fit are shown in Figure~\ref{fig:x-ray} (bottom right panel). The XRT spectrum taken on January 8 is well explained with a simple absorbed power-law model, whereas, that of January 10 is better fitted with the log-parabola model. The joint \nicer~and \nustar~spectrum from January 11 is also well explained with an absorbed log-parabola model, clearly revealing the synchrotron peak. Note that we do not use \swift-XRT data in the January 11 spectral fitting due to two reasons: (i) the fit is dominated by \nicer~and \nustar~spectra because of much better photon statistics, and (ii) after removing bad channels (using {\tt ignore bad} command in XSPEC), \swift-XRT spectrum is limited up to 5 keV, thus giving no advantage over \nicer~observation.

The broadband spectral energy distribution (SED) of \bzb~during the January 8, 10, and 11 epochs are shown in Figure~\ref{fig:sed}. The archival IR-optical spectrum reveals a bump which is likely to be originated from the host galaxy and has been noticed in many extreme blazars \citep[see, e.g.,][]{2018MNRAS.477.4257C}. The long-time averaged 4FGL SED reveals an extremely hard \gm-ray spectrum suggesting the inverse Compton peak to be located at very high energies ($>$100 GeV). Note that at this redshift the extragalactic background light attenuation is also significant \citep[][]{2011MNRAS.410.2556D,2019ApJ...882L...3P}. 

\begin{figure*}[t!]
\vbox{
\includegraphics[width=\linewidth]{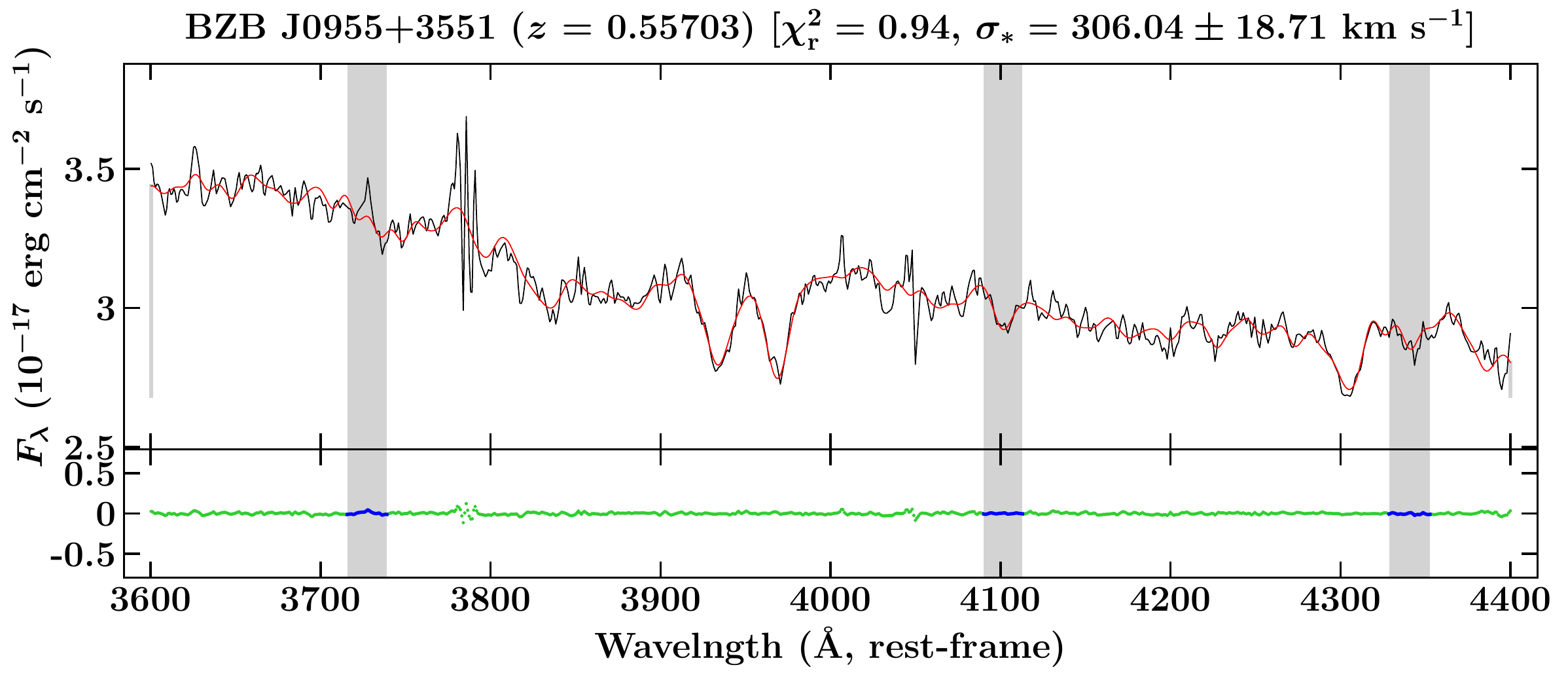} 
\includegraphics[width=\linewidth]{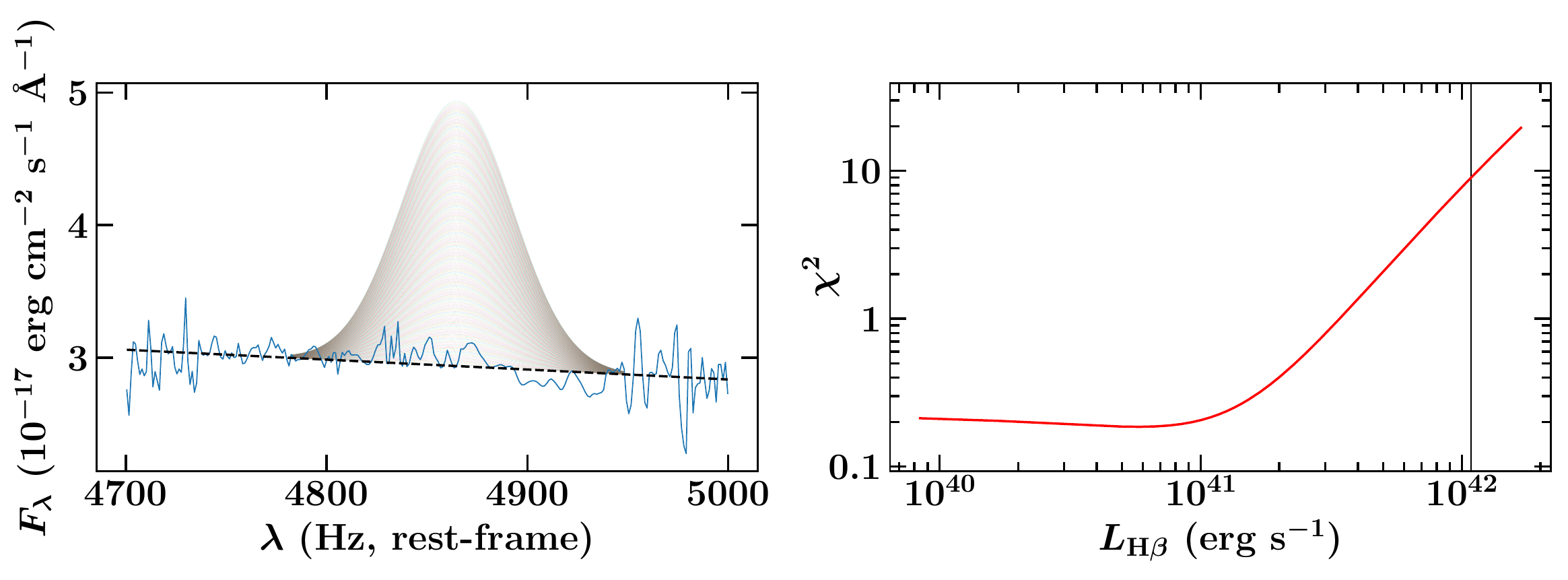} 
}
\caption{Top: Optical spectrum of BZB J0955+3551 (black line) fitted with the stellar population synthesis tool {\tt pPXF} (red line). The bottom panel refers to the residual of the fit. The grey shaded area denote the wavelength regions excluded from the fit to mask emission lines. The derived stellar velocity dispersion and the reduced $\chi^2$ are quoted. Bottom: The rest-frame OSIRIS spectrum (blue solid line, left panel) when fitted with a power-law (black dashed line) and a single Gaussian function with variable $L_{\rm H\beta}$. The right panel shows the variation of the derived $\chi^2$ as a function of $L_{\rm H\beta}$. The vertical black solid line highlights the $L_{\rm H\beta}$ value beyond which $\chi^2>\chi^2$ (99.7\%).
}
\label{fig:ppxf}
\end{figure*} 
\section{Discussion}\label{sec:discussion}
\subsection{Properties of the Central Engine}\label{subsec:engine}

We have used the well-calibrated empirical relation between the black hole mass ($M_{\rm BH}$) and the central stellar velocity dispersion ($\sigma_*$) to determine the former \citep[cf.][]{2009ApJ...698..198G,2013ARA&A..51..511K}. To determine $\sigma_{*}$, we used the penalized PiXel Fitting software \citep[{\tt pPXF};][]{2004PASP..116..138C}. This tool
works in pixel space and adopts a maximum penalized
likelihood approach to derive the line-of-sight velocity
distribution (LOSVD) from kinematic data \citep[][]{1997AJ....114..228M}. To fit the galaxy spectrum, {\tt pPXF} uses a large set of single stellar population spectral libraries which we adopted from \citet[][]{2010MNRAS.404.1639V}. It first creates a template galaxy spectrum by convolving the stellar population models with the parameterized LOSVD and then fit the model on the observed galaxy spectrum by minimizing $\chi^2$. We also added a fourth-order Legendre polynomial to account for the likely featureless contribution from the nuclear emission. From the best-fit spectrum, {\tt pPXF} computes $\sigma_*$ and associated 1$\sigma$ uncertainty. The result of this analysis is shown in Figure~\ref{fig:ppxf} and the derived $\sigma_{*}$ is $306.04\pm18.71$ km s$^{-1}$.

We used the following empirical relation to compute $M_{\rm BH}$ \citep[][]{2009ApJ...698..198G}:

\beq
\log{\left(\frac{M_{\rm BH}}{\msun}\right)} = (\msinterr) + (\msslopeerr) 
\log{\left(\frac{\sigma_*}{200\kms}\right)},
\label{e:noulmsigmafit}
\eeq

By supplying the $\sigma_*$ derived from the {\tt pPXF} fit in the above equation, the mass of the central black hole is obtained as $\log M_{\rm BH,\odot}=8.90\pm0.16$. The quoted uncertainty is statistical only and does not include the intrinsic scatter ($\sim$0.4 dex) associated with this method \citep[][]{2009ApJ...698..198G}.

Since no broad emission lines are detected in the optical spectrum of \bzb, we have determined 3$\sigma$ upper limit on the broad line region (BLR) luminosity ($L_{\rm BLR}$) by adopting the following procedure. The OSIRIS spectrum was analyzed in the rest-frame wavelength range [4700, 5000] \AA~where H$_{\beta}$ emission line is expected to be present. We brought the spectrum to the rest-frame and fitted with a power-law to reproduce the continuum. We assumed the H$\beta$ emission line as a Gaussian with variable luminosity while keeping its full width at half maximum fixed to 4000 km s$^{-1}$, a value typical for blazars \citep[cf.][]{2012ApJ...748...49S}. Then, a $\chi^2$ test was performed by fitting the Gaussian model on the data by varying the line luminosity ($L_{\rm H\beta}$). We computed the upper limit to $L_{\rm H\beta}$ when $\chi^2>\chi^2$ (99.7\%), i.e., at 3$\sigma$ confidence level. The derived upper limit on $L_{\rm H\beta}$ is $1.1\times10^{42}$ \lum. This is demonstrated in the bottom panel of Figure~\ref{fig:ppxf}. Furthermore, by adopting the line flux ratios from \citet[][]{1991ApJ...373..465F} and \citet[][]{1997MNRAS.286..415C}, we estimated the $L_{\rm BLR}$ upper limit as $\sim$2.7$\times$10$^{43}$ \lum. The presence of a more luminous BLR can be ruled out as that would emit stronger emission lines which should be observed in the optical spectrum. Furthermore, the inferred $L_{\rm BLR}$ implies an accretion rate (in Eddington units) of $L_{\rm BLR}/L_{\rm Edd}\lesssim 0.0003$. Such a low-accretion rate suggests a radiatively inefficient accretion process and is expected in BL Lac objects.

\subsection{General theoretical considerations}
\label{sec:general}

The following section considers general energetic requirements for the production of a detectable
IceCube neutrino flux in the jet of BZB J0955+3551. These are constrained by the 
observed UV -- X-ray flux just after the detection of the neutrino event, on 2020 January 8 and 
are similar to that reported for neutrino production in TXS~0506+056 by \cite{Reimer19}. 
Specifically a flux around $\sim 10^{16}$ -- $10^{17}$~Hz of $\nu F_{\nu}^{UV-X} \sim 10^{-12} \, 
F_{UV, -12}$~erg~cm$^{-2}$~s$^{-1}$ was observed, while the peak of the 
X-ray spectrum was located around $\sim 10^{18}$~Hz at a flux close to 
$\nu F_{\nu} \sim 3 \times 10^{-12} $~erg~cm$^{-2}$~s$^{-1} \equiv 10^{-12} \, 
F_{X, -12}$~erg~cm$^{-2}$~s$^{-1}$ with $F_{X, -12} \sim 3$.
We first derive a general constraint on the jet content of 
protons that might potentially be responsible for very high-energy (VHE) neutrino production, and then consider two
possibilities for the source of target photons for photo-pion production on those protons. 

The neutrino emission region propagates along the jet with Lorentz factor $\Gamma = 10 \,
\Gamma_1$, leading to Doppler boosting characterized by a Doppler factor $D = 10 \, D_1$. 
The observed sub-hour-scale X-ray variability suggests a size of the X-ray emission region of $R_X \lesssim 3.5 \times 10^{14} \, D_1$~cm. As our analytical and numerical modeling results below will demonstrate, it is unlikely that the observed X-ray emission has been produced in the same (photo-hadronic and cascade) processes as the neutrino emission. Hence, the neutrino emission region region may be different from that producing X-rays. Assuming a neutrino emission region of the size mentioned above would lead to an unrealistically high compactness, with the required relativistic proton pressure exceeding the magnetic pressure by many orders of magnitude. 
For example, the assumed production of the observed neutrino flux through photo-pion processes requires characteristic proton powers of $L_{\rm p}\sim 10^{48}-10^{49}$ \lum. Writing $L_{\rm p} = 10^{48} L_{48}$ \lum, the energy density in relativistic protons is then $u'_{\rm p} \sim 10^3 L_{48} / (R_{\rm 16}^2$) erg~cm$^{-3}$ assuming a proton escape time scale of $t'_{\rm esc} = 10^7$ sec \citep[see][for details]{2013ApJ...768...54B}. Assuming relativistic protons, the pressure exerted by protons us $p'_{\rm p} \sim u'_{\rm p}/3 \sim 350 L_{48} / (R_{16}^2$) dyne~cm$^{-2}$. The magnetic pressure, on the other hand, is $p'_{\rm B} \sim 400 B_2^2$ dyne~cm$^{-2}$. Thus, for an emission region size $\ll$10$^{16}$ cm, the proton pressure will exceed the magnetic pressure for any plausible value of the magnetic field. Thus, confinement of the emission region in such a small volume appears implausible. We therefore assume that neutrinos are produced in a larger emission region of size $R \sim 10^{16} \, R_{16}$~cm.
In the following, primes denote quantities in the rest-frame of this emission region. The 
redshift of $z = 0.5573$ corresponds to a luminosity distance of $d_L \sim 3.2 \, {\rm Gpc}
\sim 9.7 \times 10^{27}$~cm.

\subsection{Proton-photon interactions and neutrino production}
\label{sec:NeutrinoProduction}

In AGN jets, neutrinos are most plausibly produced through photo-hadronic interactions of 
relativistic protons of energy $E'_p = \gamma_p \, m_p c^2$ with target photons of energy $E'_t$. This interaction is most
efficient when the center-of-momentum frame energy squared, $s = (m_p c^2)^2 + 2 \, E'_p \, E'_t \, 
(1 - \beta_p \cos\theta)$ --- where $\beta_p = \sqrt{1 - 1 / \gamma_p^2}$ is the normalized 
velocity (in units of the speed of light c) and $\gamma_p$ the Lorentz factor of the proton --- 
of the interaction is near the $\Delta^+$ resonance, $s \sim E_{\Delta^+}^2 = (1232 {\rm MeV})^2$,
where the $p\gamma$ interaction cross section peaks. 
This translates into a condition $E'_p \, E'_t \sim 3.2 \times 10^5 \, {\rm MeV}^2$. 

The proton energy required to produce neutrinos at observed (i.e., Doppler-boosted) energies 
of hundreds of TeV, $E_{\nu} \equiv 100 \, E_{14}$~TeV ($E_{14}$ is the neutrino energy in units of $10^{14}$~eV) is $E'_p \simeq 200 \, E_{14} / (D_{1} \,
\xi_{0.05})$~TeV (i.e., $\gamma'_p = E'_p/m_p c^2 \simeq 2 \times 10^5 E_{14} / (D_{1} \, \xi_{0.05})$) 
where $\xi \equiv 0.05 \, \xi_{0.05}$ is the average neutrino energy per initial proton energy in 
photo-hadronic interactions \citep{Muecke99}. The Larmor radius of protons with such energy is 
$r_L \sim 6.7 \times 10^{10} \, (\gamma_p^\prime / [2 \times 10^5]) \, B_2^{-1}$~cm, where $B = 100 \, B_2$~G
is the magnetic field. This indicates that they are expected to be well confined within the emission 
region and can plausibly be accelerated by standard mechanisms.

For photo-pion (and neutrino) production by protons of this energy at the $\Delta^+$ resonance, 
target photons of $E'_t \geq 1.6 D_{1} \xi_{0.05}/E_{14}$~keV are required. In section
\ref{sec:target}, we will discuss two extreme options for the source of such target photons:
(a) the co-moving electron synchrotron radiation field, and (b) an external radiation field that
is isotropic in the AGN rest frame. First, however, we derive constraints on the number of relativistic 
protons that may be present in the jet.

\subsection{Constraints on Jet Power}
\label{sec:power}

Protons of energy $E'_p = 200 \, E_{14} / (D_1 \, \xi_{0.05})$~TeV radiate proton synchrotron 
radiation at a characteristic frequency of 

$$\nu_{\rm psy}^{\rm obs} = 4.2 \times 10^6 \, B_G \, \gamma_p^2 \, D \, (m_e / m_p) \; {\rm Hz}
$$
\begin{equation}
\approx 9.2 \times 10^{16} \, {B_2 \, E_{14}^2 \over D_1 \, \xi_{0.05}^2} \; {\rm Hz},
\label{nypsy}
\end{equation}
i.e., in soft X-rays. Given a number of protons of energy $\gamma_p$, i.e., $N_p (\gamma_p) \sim \gamma_p
\, dN_p(\gamma_p) / d\gamma_p$, one may calculate the produced co-moving luminosity in proton synchrotron 
radiation as 

$$
L'_{\rm psy} \approx {c \, \sigma_T \, B^2 \over 6 \pi} \, \left( {m_e \over m_p} \right)^2 \, \gamma_p^3
\, {dN_p (\gamma_p) \over d\gamma_p} 
$$
\begin{equation}
\approx 2.5 \times 10^{-2} \, {dN_p (\gamma_p) \over d\gamma_p} \,
B_2^2 \, {E_{14}^3 \over D_1^3 \, \xi_{0.05}^3} \; {\rm erg \, s}^{-1}. 
\label{Lpsy}
\end{equation}

The resulting observable soft X-ray flux, $\nu F_{\nu}^{\rm psy} \sim D_1^4 \, L'_{\rm psy} / (4 \, \pi \,
d_L^2)$ may not over-shoot the actually observed UV -- soft X-ray flux, thus constraining the differential 
number of protons to 

\begin{equation}
{dN_p (\gamma_p) \over d\gamma_p} \lesssim 4.7 \times 10^{42} \, {F_{X, -12} \, \xi_{0.05}^3 \over
B_2^2 \, D_1 \, E_{14}^3}. 
\label{dNdgammap}
\end{equation}

We consider a proton spectrum of the form $N_p (\gamma_p) = N_0 \, \gamma_p^{-\alpha_p}$ with 
$\alpha_p = 2$, extending from $\gamma_{\rm p, min} = 1$ to $\gamma_{\rm p, max} \sim 2 \times 10^5
\, E_{14} / (D_1 \, \xi_{0.05})$ so that the resulting proton synchrotron spectrum actually peaks 
(in $\nu F_{\nu}$ representation) at the characteristic proton synchrotron frequency $\nu_{\rm psy}$
evaluated in Eq. (\ref{nypsy}) above. The proton-spectrum normalization is then constrained to

\begin{equation}
N_0 \lesssim 1.9 \times 10^{53} \, {F_{X, -12} \, \xi_{0.05} \over B_2^2 \, D_1^3 \, E_{14}}
\label{N0}
\end{equation}
limiting the kinetic jet power in relativistic protons to

$$ L_p \sim {3 \, \Gamma^2 \, c \, (m_p c^2) \, N_0 \over 4 \, R} \, \ln\gamma_{\rm p,max}
$$
\begin{equation}
\lesssim 6.4 \times 10^{46} \, {F_{X, -12} \, \xi_{0.05} \, \Gamma_1^2 \over B_2^2 \, D_1^3 \,
E_{14} \, R_{16}} \, (12.2 + \eta) \; {\rm erg \, s}^{-1},
\label{Lp}
\end{equation}
where $\eta \equiv \ln (E_{14} / [D_1 \, \xi_{0.05})$. For $\eta = 0$ and all baseline parameters, 
Eq. (\ref{Lp}) evaluates to $2.3 \times 10^{48}$~erg~s$^{-1}$ for one-sided jet. In the following, we ignore the 
(presumed small) correction arising from potential values of $\eta \ne 0$. 

Using the proton spectrum with normalization given by Eq. (\ref{N0}), we may estimate the co-moving 
neutrino luminosity through the proton energy loss rate due to photopion production, given by \citet{KA08} \citep[see also][]{1988A&A...199....1B,2000PhRvD..62i3005S}

\begin{equation}
{\dot\gamma}_{\rm p, p\gamma} \approx - c \, \langle \sigma_{p\gamma} f \rangle \, n'_{\rm ph} 
(\epsilon'_t) \, \epsilon'_t \, \gamma_p
\label{gpdot}
\end{equation}

where $\epsilon'_t = E'_t / (m_e c^2)$ and $\langle \sigma_{p\gamma} f \rangle \approx 10^{-28}$~cm$^2$ 
is the elasticity-weighted p$\gamma$ interaction cross section. The factor $n'_{\rm ph} (\epsilon'_t) \, 
\epsilon'_t$ provides a proxy for the co-moving energy density of the target photon field, $u'_{\rm t} 
\approx m_e c^2 \, n'_{\rm ph} (\epsilon'_t) \, (\epsilon'_t)^2$. Considering that the energy lost by 
protons in p$\gamma$ interactions is shared approximately equally between photons and neutrinos, the VHE
neutrino luminosity is given by

$$
L'_{\nu} \approx {1 \over 2} \,  N_0 \, m_p c^2 \, \int\limits_{\gamma_1}^{\gamma_{\rm p,max}} \gamma_p^{-2} \, 
\left\vert {\dot\gamma}_{\rm p, p\gamma} \right\vert \, d\gamma_p 
$$
\begin{equation}
\approx {1 \over 2} \, c \, N_0 \, m_p c^2 \, \langle \sigma_{p\gamma} f \rangle \, {u'_t \over \epsilon'_t \, 
m_e c^2} \, \ln\left( {\gamma_{\rm p, max} \over \gamma_1} \right) 
\label{Lnu1}
\end{equation}

Considering that the target photon is unlikely to be mono-energetic, we set the lower limit in the 
integral in Eq. (\ref{Lnu1}) to $\gamma_1 = 6.4 \times 10^4 / (D_1 \, \xi_{0.05})$, corresponding 
to protons producing neutrinos with an observed energy of $\sim 30$~TeV. With this choice, the limit 
on the neutrino luminosity (corresponding to the limit on $N_0$ from Equation \ref{N0}) evaluates to 

\begin{equation}
L'_{\nu} \lesssim 1.7 \times 10^{41} \, {u'_t \over {\rm erg \, cm}^{-3}} \ln (3.1 \, E_{14}) \,
{F_{X, -12} \, \over B_2^2 \, D_1^4} \; {\rm erg \, s}^{-1},
\label{Lnu2}
\end{equation}
which yields a limit on the VHE neutrino flux measured on Earth as

\begin{equation}
F_{\nu} \lesssim 1.4 \times 10^{-12} \, {u'_t \over {\rm erg \, cm}^{-3}} \, \ln (3.1 \, E_{14}) 
\, {F_{X, -12} \over B_2^2} \; {\rm erg \, cm}^{-2} \, {\rm s}^{-1}.
\label{Fnu}
\end{equation}

The single neutrino detection of IC-200107A corresponds to an approximate neutrino energy flux of 
$F_{\nu}^{\rm obs} \sim 10^{-11} $~erg~cm$^{-2}$~s$^{-1}$ (Section~\ref{sec:nu_fl}).
Accounting for a possible Eddington bias due to the large number of potentially similar 
blazars from which no neutrinos have been detected \citep{Strotjohann19}, the actual neutrino
flux from this individual source may, however, be up to a factor of $\sim 100$ lower than 
the estimate provided above. We, therefore, base our estimates below on a neutrino flux of $F_{\nu} \sim 10^{-13} \, F_{\nu, -13}$~erg~cm$^{-2}$~s$^{-1}$. 
Thus, Eq. (\ref{Fnu}) translates into a limit on the co-moving target photon field energy 
density of

\begin{equation} 
u'_t \gtrsim 6.8 \times 10^{-2} \, {B_2^2 \, F_{\nu, -13} \over F_{X, -12} \, \ln(3.1 \, E_{14})} \; 
{\rm erg \, cm}^{-3}.  
\label{ut_required}
\end{equation}

In the following, we will 
discuss implications for the nature of such a target photon field.

\subsection{Implications on the target photon fields}
\label{sec:target}

We will distinguish two possible scenarios, which can be thought of as extreme, limiting cases: 
(a) a target photon field that is co-moving with the emission region (such as the electron-synchrotron 
emission), or (b) a stationary target photon field in the AGN rest frame.

\subsubsection{a) Co-moving target photon field}

If the target photon field is co-moving (e.g., the electron-synchrotron photon field, which is 
routinely used in lepto-hadronic blazar models as targets for photo-pion production), the target 
photon energy $E'_t$ corresponds to an observed photon energy of $E_{\rm t, obs} \geq 16 \, D_{1}^2 
\xi_{0.05}/E_{14}$~keV (i.e., hard X-rays). The directly observed X-ray flux corresponding to 
the co-moving radiation energy density from Eq. (\ref{ut_required}) is, in this case, Doppler
boosted by a factor of $D^4$ with respect to the observer, yielding a lower limit on the X-ray 
flux from this radiation field of

$$
F_t^{\rm obs, a} \sim {D^4 \, R^2 \, c \, u'_t \over d_L^2}
$$
\begin{equation}
\gtrsim 2.2 \times 10^{-11} \, { F_{\nu, -13} \, R_{16}^2 \, B_2^2 \, D_1^4 \over 
F_{X, -12} \, \ln (3.1 \, E_{14})} \; {\rm erg \, cm}^{-2} \, {\rm s}^{-1}.  
\label{Ft_a}
\end{equation}
Thus, a slightly smaller emission region size than $10^{16}$~cm, a magnetic field of $B < 10^2$~G, and/or a slightly smaller Doppler factor, $D < 10$ might plausibly allow this estimate not to over-predict the observed X-ray flux of $F_{X} \sim 3 \times 10^{-12}$~erg~cm$^{-2}$~s$^{-1}$. 
This is contrary to the case of TXS 0506+056, where the co-moving electron synchrotron radiation field being the dominant 
target photon field for photo-pion production to produce a significant flux of VHE neutrinos could be safely ruled out
\citep[see, e.g.,][]{2018ApJ...864...84K, Reimer19,2019ApJ...874L..29R,2020ApJ...889..118Z}.

\subsubsection{b) Stationary photon field in the AGN rest frame} 

In the case of a photon field that is stationary (and quasi-isotropic) in the AGN rest frame, the 
external (stationary) target photon field is Doppler boosted into the blob frame, so that 
$E_{\rm t, obs} \geq 0.16 \xi_{0.05}/E_{14}$~keV (i.e., UV-to-soft X-rays). In this case, the 
target photon density is enhanced in the co-moving frame, compared to the AGN-rest-frame energy 
density $u_t^{\rm AGN} \sim u'_t/\Gamma^2$. Assuming that the target photon field originates in
a larger region of size $R_t = 10^{17} \, R_{t, 17}$~cm surrounding the jet, the resulting,
directly observable UV -- soft X-ray flux amounts to

$$
F_t^{\rm obs, b} \sim {u'_t \, R_t^2 \, c \over \Gamma^2 \, d_L^2}
$$
\begin{equation}
\gtrsim 2.2 \times 10^{-15} \, {F_{\nu, -13} \, R_{t, 17}^2 \, B_2^2 \over \Gamma_1^2 \, F_{X, -12} 
\, \ln(3.1 \, E_{14})} \; {\rm erg \, cm}^{-2} \, {\rm s}^{-1}.
\label{Ft_b}
\end{equation}
For most plausible parameter choices this remains of the order of the observed X-ray flux. 

We therefore conclude that a scenario involving a dominant external radiation field as the target for photo-hadronic neutrino production in the jet 
of BZB J0955+3551 can more easily satisfy all observational constraints. However, a co-moving 
target photon field (e.g., electron-synchrotron) can not be ruled out or even strongly
disfavoured.

Using the 3$\sigma$ upper limit on $L_{\rm BLR}$ derived in Section~\ref{subsec:engine}, we can determine whether the external photon field needed for neutrino production (Equation~\ref{ut_required}) could originate from the BLR of \bzb. Assuming the emission region being located within a spherical BLR of radius $R_{\rm BLR}$, the comoving frame BLR energy density can be written as follows:

\begin{equation}
    u'_{\rm BLR} \sim \frac{\Gamma^2 L_{\rm BLR}}{4\pi R^2_{\rm BLR}c}
\end{equation}
Adopting $R_{\rm BLR}=R_t \sim 10^{17}$ cm and $\Gamma=10$, we get $u'_{\rm BLR}\lesssim 0.72$ erg cm$^{-3}$. This implies that the BLR could act as a reservoir of seed photons for photo-hadronic neutrino production. However, since a definite value of $L_{\rm BLR}$ could not be ascertained, a strong conclusion cannot be made. Indeed, if we consider the disk luminosity-BLR radius relationship \citep[see e.g.,][]{2008MNRAS.386..945T}, a low level of accretion luminosity of \bzb~indicates a small $R_{\rm BLR}$ and, on average, the emission region would be located farther away, as also reported in various blazar population studies \citep[see, e.g.,][]{2017ApJ...851...33P,2019ApJ...881..154P}. In the absence of a strong photon field, relativistic electrons can reach up to very high energies leading to the observation of a high synchrotron peaked SED \citep[cf.][]{2013MNRAS.432L..66G}. Therefore, though plausible, a BLR origin of the external photon field cannot be ascertained with high confidence.

In the following, we will further investigate the possibility of photo-hadronic neutrino 
production in BZB J0955+3551 with an external target photon field, and check whether the observed 
optical -- UV -- X-ray spectrum is consistent with constraints from electromagnetic cascades 
initiated by the neutrino-producing pion and muon decay processes. 

\subsection{Numerical Simulations}
Following the analytical considerations in the previous sub-sections, we now attempt to reproduce the 
observed neutrino flux with a detailed numerical model while not overshooting the observed emission.
We employ the steady-state, single-zone lepto-hadronic model described in \citet[][]{2013ApJ...768...54B}, using 
parameters 
in agreement with the limits derived from the analytical estimates in the previous sub-section. 
In the numerical simulation, as described in \citet[][]{2013ApJ...768...54B}, the code determines the radiating proton spectrum by evaluating an equilibrium between injection of a power-law proton spectrum, escape, and radiative cooling. The escape time scale has been set as a multiple $\eta_{\rm esc} = 30$ times the light crossing time scale, i.e., $t'_{\rm esc} = \eta_{\rm esc} * R / c$ (in the co-moving frame of the emission region). This is the same for electrons and protons. Numerically, therefore, a proton escape time scale of $t'_{\rm esc} = 10^7$ sec was used in the simulation. The external photon field required
for photo-hadronic neutrino production is represented by an equivalent electron-synchrotron radiation field
with the same characteristics 
as the presumed external radiation field in the
co-moving frame of the neutrino emission region,
because the code of \citet[][]{2013ApJ...768...54B} does currently
not include external radiation fields for photo-pion production. It has been shown that
the anisotropy of the target photon field has a negligible effect on the neutrino production and electromagnetic
radiation output\footnote{\url{ https://indico.cern.ch/event/828038/contributions/3590902/}}. Thus, our equivalent electron-synchrotron radiation set-up is an appropriate proxy for the
required external target photon field. 

The results of the simulation reproducing the 
Eddington-bias-corrected neutrino flux of $F_{\nu} \sim 10^{-13}$~erg~cm$^{-2}$~s$^{-1}$ 
are shown in Figure~\ref{fig:sed}
with red and violet solid lines. The corresponding parameters are listed in Table~\ref{tab:sed_param}. The choice of the SED parameters was based on the typical values found in previous lepto-hadronic modeling of blazars \citep[][]{2013ApJ...768...54B,2018ApJ...864...84K, Reimer19,2019ApJ...874L..29R}. The very intense target
photon field provides a very high \gm\gm~opacity for \gm-rays in the \fermi-LAT energy range and higher, analogous
to what was found for TXS 0506+056 \citep[][]{Reimer19}. This suggests that one would not expect a significant
correlation between neutrino and \gm-ray activity. 
Furthermore, the cascade synchrotron flux is well below the 
observed optical -- UV -- X-ray flux, suggesting that all
electromagnetic flux components are likely to be produced
by different processes and possibly even in a different
emission region than the neutrino flux.

The dashed model curves in Figure~\ref{fig:sed} show
an attempt to reproduce the neutrino flux 
of $10^{-11}$ \ergflux,
i.e., neglecting the Eddington bias. For
this purpose, the target photon density was increased by a factor of 36 with respect to the previous simulation, leaving
all other parameters unchanged. It is obvious that, in this case, the electromagnetic output from proton synchrotron
and cascades overshoots the 
optical -- UV fluxes, and its spectral shape is
very different from the observed optical -- X-ray spectrum
of BZB J0955+3551. This can, therefore, be
ruled out. 

Overall, our results suggest a scenario in which a relativistic proton population responsible for the
observed neutrino emission from \bzb~may only make a sub-dominant contribution to the observed
X-ray flare. The detection of rapid X-ray flux variability also hints that the neutrino producing region may not be same as the one emitting X-rays. Therefore, we argue that the detection of an X-ray flare from \bzb~found close-in-time to IC-200107A is likely a coincidence and the two events may not be physically connected.

A comprehensive analysis of the this neutrino event has also been carried out by \citet[][]{2020arXiv200507218P} who studied the same event using various lepto-hadronic models with different emission region conditions. Though co-spatial neutrino and electromagnetic radiation producing regions were considered, the neutrino emission was not found to be related with the observed X-ray flare. These results are aligned with our findings derived from analytical calculation and a crude numerical simulation as discussed above. Moreover, among various theoretical models, they also explored a case of hidden external photon field (a putative weak BLR) providing seed photons for photo-hadronic production of neutrinos. In this single-zone lepto-hadronic model with external photon field, they reported that the predicted neutrino flux would be even lower during the X-ray flare to avoid overshooting the observed \gm-ray spectrum \citep[see][for details]{2020arXiv200507218P}. These findings are in agreement with that reported in this work.

\begin{deluxetable}{ll}[t!]
\tablecaption{Parameters used/derived from the numerical simulation. The shape of the proton spectrum is adopted as a power law.\label{tab:sed_param}}
\tablewidth{0pt}
\tablehead{
\colhead{Parameter} & \colhead{Value}}
\startdata
Co-moving photon field energy density (erg cm$^{-3}$) & 55 \\
Co-moving photon field peak frequency (Hz) &  $3.5\times10^{17}$ \\
Magnetic field (Gauss) & 100 \\
Bulk Lorentz factor  & 10 \\
Emission-region radius (cm) & $1\times10^{16}$\\
Viewing angle (degrees) & 5.7 \\
Low-energy cut-off of proton spectrum (GeV)  & 1 \\
Proton high-energy cut-off (GeV) & 10$^6$ \\
Proton Injection spectral index  & 1.1 \\
Kinetic luminosity in protons (\lum) & $1\times10^{49}$ \\
Magnetic jet power (\lum) & $3.75\times10^{47}$ \\
\enddata
\end{deluxetable}

\section{Summary}\label{sec:summary}

We have followed the X-ray flaring activity of \bzb~\citep[][]{2020ATel13394....1G,2020ATel13395....1K} with \nustar, \swift, and GTC and also used the simultaneous observation from \nicer. Using the high-quality OSIRIS spectrum, we determined the spectroscopic redshift of the blazar as $z=0.55703^{+0.00033}_{-0.00021}$. On the other hand, we could not ascertain the nature of the companion object identified $\sim$3$^{\prime\prime}$ South-East of \bzb~in the $i'$ filter Pan-STARRS image. From the stellar velocity dispersion measured using {\tt pPXF}, the central black hole mass of \bzb~was derived as 10$^{8.90\pm0.16}$~\msun. Moreover, the optical spectrum of the source reveals a faint [O II]3727 emission line with rest-frame equivalent width of 0.15$\pm$0.05 \AA. We estimated a very low-level of accretion activity which is consistent with that expected from BL Lac objects. There are tentative evidences ($\lesssim3.5\sigma$) for the hour-scale flux variability in the X-ray band, as estimated from the \nustar~and \nicer~light curves. Finally, we showed that 
a scenario involving an external photon field as 
targets for photo-pion production of neutrinos is more
easily able to satisfy all observational constraints 
but a scenario invoking a co-moving target photon field (e.g., electron-synchrotron) can not be
ruled out or even strongly disfavored. Any electromagnetic signatures of the photo-pion processes
responsible for the neutrino emission, are likely to 
only make a sub-dominant contribution to the observed
electromagnetic radiation from IR to $\gamma$-rays suggesting that the X-ray flaring event may not be directly connected with IC-200107A.

\bibliography{Master.bib}{}

\begin{thebibliography}{}
\expandafter\ifx\csname natexlab\endcsname\relax\def\natexlab#1{#1}\fi
\providecommand{\url}[1]{\href{#1}{#1}}
\providecommand{\dodoi}[1]{doi:~\href{http://doi.org/#1}{\nolinkurl{#1}}}
\providecommand{\doeprint}[1]{\href{http://ascl.net/#1}{\nolinkurl{http://ascl.net/#1}}}
\providecommand{\doarXiv}[1]{\href{https://arxiv.org/abs/#1}{\nolinkurl{https://arxiv.org/abs/#1}}}

\bibitem[{{Aartsen} {et~al.}(2015){Aartsen}, {Ackermann}, {Adams}, {Aguilar},
  {Ahlers}, {Ahrens}, {Altmann}, {Anderson}, {Archinger}, {Arguelles}, {Arlen},
  {Auffenberg}, {Bai}, {Barwick}, {Baum}, {Bay}, {Baker}, {Beatty}, {Becker
  Tjus}, {Becker}, {BenZvi}, {Berghaus}, {Berley}, {Bernardini}, {Bernhard},
  {Besson}, {Binder}, {Bindig}, {Bissok}, {Blaufuss}, {Blumenthal}, {Boersma},
  {Bohm}, {Bos}, {Bose}, {B{\"o}ser}, {Botner}, {Brayeur}, {Bretz}, {Brown},
  {Buzinsky}, {Casey}, {Casier}, {Cheung}, {Chirkin}, {Christov}, {Christy},
  {Clark}, {Classen}, {Clevermann}, {Coenders}, {Cowen}, {Cruz Silva},
  {Daughhetee}, {Davis}, {Day}, {de Andr{\'e}}, {De Clercq}, {Dembinski}, {De
  Ridder}, {Desiati}, {de Vries}, {de Wasseige}, {de With}, {DeYoung},
  {D{\'\i}az-V{\'e}lez}, {Dumm}, {Dunkman}, {Eagan}, {Eberhardt}, {Ehrhardt},
  {Eichmann}, {Eisch}, {Euler}, {Evenson}, {Fadiran}, {Fazely}, {Fedynitch},
  {Feintzeig}, {Felde}, {Filimonov}, {Finley}, {Fischer-Wasels}, {Flis},
  {Frantzen}, {Fuchs}, {Gaisser}, {Gaior}, {Gallagher}, {Gerhardt}, {Gier},
  {Gladstone}, {Gl{\"u}senkamp}, {Goldschmidt}, {Golup}, {Gonzalez}, {Goodman},
  {G{\'o}ra}, {Grant}, {Gretskov}, {Groh}, {Gro{\ss}}, {Ha}, {Haack}, {Haj
  Ismail}, {Hallen}, {Hallgren}, {Halzen}, {Hanson}, {Hebecker}, {Heereman},
  {Heinen}, {Helbing}, {Hellauer}, {Hellwig}, {Hickford}, {Hignight}, {Hill},
  {Hoffman}, {Hoffmann}, {Homeier}, {Hoshina}, {Huang}, {Huelsnitz}, {Hulth},
  {Hultqvist}, {In}, {Ishihara}, {Jacobi}, {Jacobsen}, {Japaridze}, {Jero},
  {Jurkovic}, {Kaminsky}, {Kappes}, {Karg}, {Karle}, {Kauer}, {Keivani},
  {Kelley}, {Kheirandish}, {Kiryluk}, {Kl{\"a}s}, {Klein}, {K{\"o}hne},
  {Kohnen}, {Kolanoski}, {Koob}, {K{\"o}pke}, {Kopper}, {Kopper}, {Koskinen},
  {Kowalski}, {Krings}, {Kroll}, {Kroll}, {Kunnen}, {Kurahashi}, {Kuwabara},
  {Labare}, {Lanfranchi}, {Larsen}, {Larson}, {Lesiak-Bzdak}, {Leuermann},
  {L{\"u}nemann}, {Madsen}, {Maggi}, {Mahn}, {Maruyama}, {Mase}, {Matis},
  {Maunu}, {McNally}, {Meagher}, {Medici}, {Meli}, {Meures}, {Miarecki},
  {Middell}, {Middlemas}, {Milke}, {Miller}, {Mohrmann}, {Montaruli}, {Morse},
  {Nahnhauer}, {Naumann}, {Niederhausen}, {Nowicki}, {Nygren}, {Obertacke},
  {Olivas}, {Omairat}, {O'Murchadha}, {Palczewski}, {Paul}, {Pepper},
  {P{\'e}rez de los Heros}, {Pfendner}, {Pieloth}, {Pinat}, {Posselt}, {Price},
  {Przybylski}, {P{\"u}tz}, {Quinnan}, {R{\"a}del}, {Rameez}, {Rawlins},
  {Redl}, {Rees}, {Reimann}, {Relich}, {Resconi}, {Rhode}, {Richman}, {Riedel},
  {Robertson}, {Rodrigues}, {Rongen}, {Rott}, {Ruhe}, {Ruzybayev}, {Ryckbosch},
  {Saba}, {Sand er}, {Sandroos}, {Santander}, {Sarkar}, {Schatto}, {Scheriau},
  {Schmidt}, {Schmitz}, {Schoenen}, {Sch{\"o}neberg}, {Sch{\"o}nwald},
  {Schukraft}, {Schulte}, {Schulz}, {Seckel}, {Sestayo}, {Seunarine},
  {Shanidze}, {Smith}, {Soldin}, {Spiczak}, {Spiering}, {Stamatikos}, {Stanev},
  {Stanisha}, {Stasik}, {Stezelberger}, {Stokstad}, {St{\"o}{\ss}l},
  {Strahler}, {Str{\"o}m}, {Strotjohann}, {Sullivan}, {Sutherland}, {Taavola},
  {Taboada}, {Tamburro}, {Ter-Antonyan}, {Terliuk}, {Te{\v{s}}i{\'c}}, {Tilav},
  {Toale}, {Tobin}, {Tosi}, {Tselengidou}, {Unger}, {Usner}, {Vallecorsa}, {van
  Eijndhoven}, {Vand enbroucke}, {van Santen}, {Vanheule}, {Vehring}, {Voge},
  {Vraeghe}, {Walck}, {Wallraff}, {Weaver}, {Wellons}, {Wendt}, {Westerhoff},
  {Whelan}, {Whitehorn}, {Wichary}, {Wiebe}, {Wiebusch}, {Williams}, {Wissing},
  {Wolf}, {Wood}, {Woschnagg}, {Xu}, {Xu}, {Xu}, {Yanez}, {Yodh}, {Yoshida},
  {Zarzhitsky}, {Ziemann}, {Zoll}, \& {IceCube
  Collaboration}}]{2015ApJ...807...46A}
{Aartsen}, M.~G., {Ackermann}, M., {Adams}, J., {et~al.} 2015, \apj, 807, 46,
  \dodoi{10.1088/0004-637X/807/1/46}

\bibitem[{{Aartsen} {et~al.}(2017{\natexlab{a}}){Aartsen}, {Abraham},
  {Ackermann}, {Adams}, {Aguilar}, {Ahlers}, {Ahrens}, {Altmann}, {Andeen},
  {Anderson}, {Ansseau}, {Anton}, {Archinger}, {Arg{\"u}elles}, {Auffenberg},
  {Axani}, {Bai}, {Barwick}, {Baum}, {Bay}, {Beatty}, {Becker Tjus}, {Becker},
  {BenZvi}, {Berley}, {Bernardini}, {Bernhard}, {Besson}, {Binder}, {Bindig},
  {Bissok}, {Blaufuss}, {Blot}, {Bohm}, {B{\"o}rner}, {Bos}, {Bose},
  {B{\"o}ser}, {Botner}, {Braun}, {Brayeur}, {Bretz}, {Bron}, {Burgman},
  {Carver}, {Casier}, {Cheung}, {Chirkin}, {Christov}, {Clark}, {Classen},
  {Coenders}, {Collin}, {Conrad}, {Cowen}, {Cross}, {Day}, {de Andr{\'e}}, {De
  Clercq}, {del Pino Rosendo}, {Dembinski}, {De Ridder}, {Desiati}, {de Vries},
  {de Wasseige}, {de With}, {DeYoung}, {D{\'\i}az-V{\'e}lez}, {di Lorenzo},
  {Dujmovic}, {Dumm}, {Dunkman}, {Eberhardt}, {Ehrhardt}, {Eichmann}, {Eller},
  {Euler}, {Evenson}, {Fahey}, {Fazely}, {Feintzeig}, {Felde}, {Filimonov},
  {Finley}, {Flis}, {F{\"o}sig}, {Franckowiak}, {Friedman}, {Fuchs}, {Gaisser},
  {Gallagher}, {Gerhardt}, {Ghorbani}, {Giang}, {Gladstone}, {Glauch},
  {Gl{\"u}senkamp}, {Goldschmidt}, {Golup}, {Gonzalez}, {Grant}, {Griffith},
  {Haack}, {Haj Ismail}, {Hallgren}, {Halzen}, {Hansen}, {Hansmann}, {Hanson},
  {Hebecker}, {Heereman}, {Helbing}, {Hellauer}, {Hickford}, {Hignight},
  {Hill}, {Hoffman}, {Hoffmann}, {Holzapfel}, {Hoshina}, {Huang}, {Huber},
  {Hultqvist}, {In}, {Ishihara}, {Jacobi}, {Japaridze}, {Jeong}, {Jero},
  {Jones}, {Jurkovic}, {Kappes}, {Karg}, {Karle}, {Katz}, {Kauer}, {Keivani},
  {Kelley}, {Kheirandish}, {Kim}, {Kintscher}, {Kiryluk}, {Kittler}, {Klein},
  {Kohnen}, {Koirala}, {Kolanoski}, {Konietz}, {K{\"o}pke}, {Kopper}, {Kopper},
  {Koskinen}, {Kowalski}, {Krings}, {Kroll}, {Kr{\"u}ckl}, {Kr{\"u}ger},
  {Kunnen}, {Kunwar}, {Kurahashi}, {Kuwabara}, {Labare}, {Lanfranchi},
  {Larson}, {Lauber}, {Lennarz}, {Lesiak-Bzdak}, {Leuermann}, {Lu},
  {L{\"u}nemann}, {Madsen}, {Maggi}, {Mahn}, {Mancina}, {Mandelartz},
  {Maruyama}, {Mase}, {Maunu}, {McNally}, {Meagher}, {Medici}, {Meier}, {Meli},
  {Menne}, {Merino}, {Meures}, {Miarecki}, {Mohrmann}, {Montaruli}, {Moulai},
  {Nahnhauer}, {Naumann}, {Neer}, {Niederhausen}, {Nowicki}, {Nygren},
  {Obertacke Pollmann}, {Olivas}, {O'Murchadha}, {Palczewski}, {Pandya},
  {Pankova}, {Peiffer}, {Penek}, {Pepper}, {P{\'e}rez de los Heros}, {Pieloth},
  {Pinat}, {Price}, {Przybylski}, {Quinnan}, {Raab}, {R{\"a}del}, {Rameez},
  {Rawlins}, {Reimann}, {Relethford}, {Relich}, {Resconi}, {Rhode}, {Richman},
  {Riedel}, {Robertson}, {Rongen}, {Rott}, {Ruhe}, {Ryckbosch}, {Rysewyk},
  {Sabbatini}, {Sanchez Herrera}, {Sand rock}, {Sandroos}, {Sarkar},
  {Satalecka}, {Schlunder}, {Schmidt}, {Schoenen}, {Sch{\"o}neberg},
  {Schumacher}, {Seckel}, {Seunarine}, {Soldin}, {Song}, {Spiczak}, {Spiering},
  {Stanev}, {Stasik}, {Stettner}, {Steuer}, {Stezelberger}, {Stokstad},
  {St{\"o}ssl}, {Str{\"o}m}, {Strotjohann}, {Sullivan}, {Sutherland},
  {Taavola}, {Taboada}, {Tatar}, {Tenholt}, {Ter-Antonyan}, {Terliuk},
  {Te{\v{s}}i{\'c}}, {Tilav}, {Toale}, {Tobin}, {Toscano}, {Tosi},
  {Tselengidou}, {Turcati}, {Unger}, {Usner}, {Vandenbroucke}, {van
  Eijndhoven}, {Vanheule}, {van Rossem}, {van Santen}, {Veenkamp}, {Vehring},
  {Voge}, {Vogel}, {Vraeghe}, {Walck}, {Wallace}, {Wallraff}, {Wandkowsky},
  {Weaver}, {Weiss}, {Wendt}, {Westerhoff}, {Whelan}, {Wickmann}, {Wiebe},
  {Wiebusch}, {Wille}, {Williams}, {Wills}, {Wolf}, {Wood}, {Woolsey},
  {Woschnagg}, {Xu}, {Xu}, {Xu}, {Yanez}, {Yodh}, {Yoshida}, {Zoll}, \&
  {IceCube Collaboration}}]{2017ApJ...835..151A}
{Aartsen}, M.~G., {Abraham}, K., {Ackermann}, M., {et~al.} 2017{\natexlab{a}},
  \apj, 835, 151, \dodoi{10.3847/1538-4357/835/2/151}

\bibitem[{{Aartsen} {et~al.}(2017{\natexlab{b}}){Aartsen}, {Ackermann},
  {Adams}, {Aguilar}, {Ahlers}, {Ahrens}, {Altmann}, {Andeen}, {Anderson},
  {Ansseau}, {Anton}, {Archinger}, {Arg{\"u}elles}, {Auffenberg}, {Axani},
  {Bai}, {Barwick}, {Baum}, {Bay}, {Beatty}, {Becker Tjus}, {Becker}, {BenZvi},
  {Berley}, {Bernardini}, {Bernhard}, {Besson}, {Binder}, {Bindig}, {Bissok},
  {Blaufuss}, {Blot}, {Bohm}, {B{\"o}rner}, {Bos}, {Bose}, {B{\"o}ser},
  {Botner}, {Braun}, {Brayeur}, {Bretz}, {Bron}, {Burgman}, {Carver}, {Casier},
  {Cheung}, {Chirkin}, {Christov}, {Clark}, {Classen}, {Coenders}, {Collin},
  {Conrad}, {Cowen}, {Cross}, {Day}, {de Andr{\'e}}, {De Clercq}, {del Pino
  Rosendo}, {Dembinski}, {De Ridder}, {Desiati}, {de Vries}, {de Wasseige}, {de
  With}, {DeYoung}, {D{\'\i}az-V{\'e}lez}, {di Lorenzo}, {Dujmovic}, {Dumm},
  {Dunkman}, {Eberhardt}, {Ehrhardt}, {Eichmann}, {Eller}, {Euler}, {Evenson},
  {Fahey}, {Fazely}, {Feintzeig}, {Felde}, {Filimonov}, {Finley}, {Flis},
  {F{\"o}sig}, {Franckowiak}, {Friedman}, {Fuchs}, {Gaisser}, {Gallagher},
  {Gerhardt}, {Ghorbani}, {Giang}, {Gladstone}, {Glauch}, {Gl{\"u}senkamp},
  {Goldschmidt}, {Gonzalez}, {Grant}, {Griffith}, {Haack}, {Hallgren},
  {Halzen}, {Hansen}, {Hansmann}, {Hanson}, {Hebecker}, {Heereman}, {Helbing},
  {Hellauer}, {Hickford}, {Hignight}, {Hill}, {Hoffman}, {Hoffmann}, {Hoshina},
  {Huang}, {Huber}, {Hultqvist}, {In}, {Ishihara}, {Jacobi}, {Japaridze},
  {Jeong}, {Jero}, {Jones}, {Kang}, {Kappes}, {Karg}, {Karle}, {Katz}, {Kauer},
  {Keivani}, {Kelley}, {Kheirandish}, {Kim}, {Kim}, {Kintscher}, {Kiryluk},
  {Kittler}, {Klein}, {Kohnen}, {Koirala}, {Kolanoski}, {Konietz}, {K{\"o}pke},
  {Kopper}, {Kopper}, {Koskinen}, {Kowalski}, {Krings}, {Kroll}, {Kr{\"u}ckl},
  {Kr{\"u}ger}, {Kunnen}, {Kunwar}, {Kurahashi}, {Kuwabara}, {Labare},
  {Lanfranchi}, {Larson}, {Lauber}, {Lennarz}, {Lesiak-Bzdak}, {Leuermann},
  {Lu}, {L{\"u}nemann}, {Madsen}, {Maggi}, {Mahn}, {Mancina}, {Mandelartz},
  {Maruyama}, {Mase}, {Maunu}, {McNally}, {Meagher}, {Medici}, {Meier}, {Meli},
  {Menne}, {Merino}, {Meures}, {Miarecki}, {Montaruli}, {Moulai}, {Nahnhauer},
  {Naumann}, {Neer}, {Niederhausen}, {Nowicki}, {Nygren}, {Obertacke Pollmann},
  {Olivas}, {O'Murchadha}, {Palczewski}, {Pandya}, {Pankova}, {Peiffer},
  {Penek}, {Pepper}, {P{\'e}rez de los Heros}, {Pieloth}, {Pinat}, {Price},
  {Przybylski}, {Quinnan}, {Raab}, {R{\"a}del}, {Rameez}, {Rawlins}, {Reimann},
  {Relethford}, {Relich}, {Resconi}, {Rhode}, {Richman}, {Riedel}, {Robertson},
  {Rongen}, {Rott}, {Ruhe}, {Ryckbosch}, {Rysewyk}, {Sabbatini}, {Sanchez
  Herrera}, {Sandrock}, {Sandroos}, {Sarkar}, {Satalecka}, {Schlunder},
  {Schmidt}, {Schoenen}, {Sch{\"o}neberg}, {Schumacher}, {Seckel}, {Seunarine},
  {Soldin}, {Song}, {Spiczak}, {Spiering}, {Stanev}, {Stasik}, {Stettner},
  {Steuer}, {Stezelberger}, {Stokstad}, {St{\"o}{\ss}l}, {Str{\"o}m},
  {Strotjohann}, {Sullivan}, {Sutherland}, {Taavola}, {Taboada}, {Tatar},
  {Tenholt}, {Ter-Antonyan}, {Terliuk}, {Te{\v{s}}i{\'c}}, {Tilav}, {Toale},
  {Tobin}, {Toscano}, {Tosi}, {Tselengidou}, {Turcati}, {Unger}, {Usner},
  {Vandenbroucke}, {van Eijndhoven}, {Vanheule}, {van Rossem}, {van Santen},
  {Vehring}, {Voge}, {Vogel}, {Vraeghe}, {Walck}, {Wallace}, {Wallraff},
  {Wandkowsky}, {Weaver}, {Weiss}, {Wendt}, {Westerhoff}, {Whelan}, {Wickmann},
  {Wiebe}, {Wiebusch}, {Wille}, {Williams}, {Wills}, {Wolf}, {Wood}, {Woolsey},
  {Woschnagg}, {Xu}, {Xu}, {Xu}, {Yanez}, {Yodh}, {Yoshida}, \&
  {Zoll}}]{2017APh....92...30A}
{Aartsen}, M.~G., {Ackermann}, M., {Adams}, J., {et~al.} 2017{\natexlab{b}},
  Astroparticle Physics, 92, 30, \dodoi{10.1016/j.astropartphys.2017.05.002}

\bibitem[{Aartsen {et~al.}(2020)}]{Aartsen:2019fau}
Aartsen, M.~G., {et~al.} 2020, Phys. Rev. Lett., 124, 051103,
  \dodoi{10.1103/PhysRevLett.124.051103}

\bibitem[{{Abdollahi} {et~al.}(2020){Abdollahi}, {Acero}, {Ackermann},
  {Ajello}, {Atwood}, {Axelsson}, {Baldini}, {Ballet}, {Barbiellini},
  {Bastieri}, {Becerra Gonzalez}, {Bellazzini}, {Berretta}, {Bissaldi}, {Bland
  ford}, {Bloom}, {Bonino}, {Bottacini}, {Brandt}, {Bregeon}, {Bruel},
  {Buehler}, {Burnett}, {Buson}, {Cameron}, {Caputo}, {Caraveo}, {Casandjian},
  {Castro}, {Cavazzuti}, {Charles}, {Chaty}, {Chen}, {Cheung}, {Chiaro},
  {Ciprini}, {Cohen-Tanugi}, {Cominsky}, {Coronado-Bl{\'a}zquez}, {Costantin},
  {Cuoco}, {Cutini}, {D'Ammando}, {DeKlotz}, {de la Torre Luque}, {de Palma},
  {Desai}, {Digel}, {Di Lalla}, {Di Mauro}, {Di Venere}, {Dom{\'\i}nguez},
  {Dumora}, {Fana Dirirsa}, {Fegan}, {Ferrara}, {Franckowiak}, {Fukazawa},
  {Funk}, {Fusco}, {Gargano}, {Gasparrini}, {Giglietto}, {Giommi}, {Giordano},
  {Giroletti}, {Glanzman}, {Green}, {Grenier}, {Griffin}, {Grondin}, {Grove},
  {Guiriec}, {Harding}, {Hayashi}, {Hays}, {Hewitt}, {Horan},
  {J{\'o}hannesson}, {Johnson}, {Kamae}, {Kerr}, {Kocevski}, {Kovac'evic'},
  {Kuss}, {Landriu}, {Larsson}, {Latronico}, {Lemoine-Goumard}, {Li},
  {Liodakis}, {Longo}, {Loparco}, {Lott}, {Lovellette}, {Lubrano}, {Madejski},
  {Maldera}, {Malyshev}, {Manfreda}, {Marchesini}, {Marcotulli},
  {Mart{\'\i}-Devesa}, {Martin}, {Massaro}, {Mazziotta}, {McEnery}, {Mereu},
  {Meyer}, {Michelson}, {Mirabal}, {Mizuno}, {Monzani}, {Morselli},
  {Moskalenko}, {Negro}, {Nuss}, {Ojha}, {Omodei}, {Orienti}, {Orlando},
  {Ormes}, {Palatiello}, {Paliya}, {Paneque}, {Pei}, {Pe{\~n}a-Herazo},
  {Perkins}, {Persic}, {Pesce-Rollins}, {Petrosian}, {Petrov}, {Piron}, {Poon},
  {Porter}, {Principe}, {Rain{\`o}}, {Rando}, {Razzano}, {Razzaque}, {Reimer},
  {Reimer}, {Remy}, {Reposeur}, {Romani}, {Saz Parkinson}, {Schinzel},
  {Serini}, {Sgr{\`o}}, {Siskind}, {Smith}, {Spandre}, {Spinelli}, {Strong},
  {Suson}, {Tajima}, {Takahashi}, {Tak}, {Thayer}, {Thompson}, {Tibaldo},
  {Torres}, {Torresi}, {Valverde}, {Van Klaveren}, {van Zyl}, {Wood},
  {Yassine}, \& {Zaharijas}}]{2020ApJS..247...33A}
{Abdollahi}, S., {Acero}, F., {Ackermann}, M., {et~al.} 2020, \apjs, 247, 33,
  \dodoi{10.3847/1538-4365/ab6bcb}

\bibitem[{{Arnaud}(1996)}]{Arnaud96}
{Arnaud}, K.~A. 1996, in Astronomical Society of the Pacific Conference Series,
  Vol. 101, Astronomical Data Analysis Software and Systems V, ed. G.~H.
  {Jacoby} \& J.~{Barnes}, 17

\bibitem[{{Berezinskii} \& {Grigor'eva}(1988)}]{1988A&A...199....1B}
{Berezinskii}, V.~S., \& {Grigor'eva}, S.~I. 1988, \aap, 199, 1

\bibitem[{{Blaufuss} {et~al.}(2019){Blaufuss}, {Kintscher}, {Lu}, \&
  {Tung}}]{2019ICRC...36.1021B}
{Blaufuss}, E., {Kintscher}, T., {Lu}, L., \& {Tung}, C.~F. 2019, in
  International Cosmic Ray Conference, Vol.~36, 36th International Cosmic Ray
  Conference (ICRC2019), 1021.
\newblock \doarXiv{1908.04884}

\bibitem[{{B{\"o}ttcher} {et~al.}(2013){B{\"o}ttcher}, {Reimer}, {Sweeney}, \&
  {Prakash}}]{2013ApJ...768...54B}
{B{\"o}ttcher}, M., {Reimer}, A., {Sweeney}, K., \& {Prakash}, A. 2013, \apj,
  768, 54, \dodoi{10.1088/0004-637X/768/1/54}

\bibitem[{{Breeveld} {et~al.}(2011){Breeveld}, {Landsman}, {Holland}, {Roming},
  {Kuin}, \& {Page}}]{2011AIPC.1358..373B}
{Breeveld}, A.~A., {Landsman}, W., {Holland}, S.~T., {et~al.} 2011, in American
  Institute of Physics Conference Series, Vol. 1358, American Institute of
  Physics Conference Series, ed. J.~E. {McEnery}, J.~L. {Racusin}, \&
  N.~{Gehrels}, 373--376, \dodoi{10.1063/1.3621807}

\bibitem[{{Cappellari} \& {Emsellem}(2004)}]{2004PASP..116..138C}
{Cappellari}, M., \& {Emsellem}, E. 2004, \pasp, 116, 138,
  \dodoi{10.1086/381875}

\bibitem[{{Celotti} {et~al.}(1997){Celotti}, {Padovani}, \&
  {Ghisellini}}]{1997MNRAS.286..415C}
{Celotti}, A., {Padovani}, P., \& {Ghisellini}, G. 1997, \mnras, 286, 415

\bibitem[{{Cepa} {et~al.}(2000){Cepa}, {Aguiar}, {Escalera},
  {Gonzalez-Serrano}, {Joven-Alvarez}, {Peraza}, {Rasilla}, {Rodriguez-Ramos},
  {Gonzalez}, {Cobos Duenas}, {Sanchez}, {Tejada}, {Bland-Hawthorn},
  {Militello}, \& {Rosa}}]{2000SPIE.4008..623C}
{Cepa}, J., {Aguiar}, M., {Escalera}, V.~G., {et~al.} 2000, in \procspie, Vol.
  4008, Optical and IR Telescope Instrumentation and Detectors, ed. M.~{Iye} \&
  A.~F. {Moorwood}, 623--631, \dodoi{10.1117/12.395520}

\bibitem[{{Cepa} {et~al.}(2003){Cepa}, {Aguiar-Gonzalez}, {Bland-Hawthorn},
  {Castaneda}, {Cobos}, {Correa}, {Espejo}, {Fragoso-Lopez}, {Fuentes},
  {Gigante}, {Gonzalez}, {Gonzalez-Escalera}, {Gonzalez-Serrano},
  {Joven-Alvarez}, {Lopez-Ruiz}, {Militello}, {Cano}, {Perez}, {Perez},
  {Rasilla}, {Sanchez}, \& {Tejada}}]{2003SPIE.4841.1739C}
{Cepa}, J., {Aguiar-Gonzalez}, M., {Bland-Hawthorn}, J., {et~al.} 2003, in
  \procspie, Vol. 4841, Instrument Design and Performance for Optical/Infrared
  Ground-based Telescopes, ed. M.~{Iye} \& A.~F.~M. {Moorwood}, 1739--1749,
  \dodoi{10.1117/12.460913}

\bibitem[{{Chang} {et~al.}(2019){Chang}, {Arsioli}, {Giommi}, {Padovani}, \&
  {Brandt}}]{2019A&A...632A..77C}
{Chang}, Y.~L., {Arsioli}, B., {Giommi}, P., {Padovani}, P., \& {Brandt}, C.~H.
  2019, \aap, 632, A77, \dodoi{10.1051/0004-6361/201834526}

\bibitem[{{Costamante} {et~al.}(2018){Costamante}, {Bonnoli}, {Tavecchio},
  {Ghisellini}, {Tagliaferri}, \& {Khangulyan}}]{2018MNRAS.477.4257C}
{Costamante}, L., {Bonnoli}, G., {Tavecchio}, F., {et~al.} 2018, \mnras, 477,
  4257, \dodoi{10.1093/mnras/sty857}

\bibitem[{{Costamante} {et~al.}(2001){Costamante}, {Ghisellini}, {Giommi},
  {Tagliaferri}, {Celotti}, {Chiaberge}, {Fossati}, {Maraschi}, {Tavecchio},
  {Treves}, \& {Wolter}}]{2001A&A...371..512C}
{Costamante}, L., {Ghisellini}, G., {Giommi}, P., {et~al.} 2001, \aap, 371,
  512, \dodoi{10.1051/0004-6361:20010412}

\bibitem[{{Dom{\'\i}nguez} {et~al.}(2011){Dom{\'\i}nguez}, {Primack},
  {Rosario}, {Prada}, {Gilmore}, {Faber}, {Koo}, {Somerville},
  {P{\'e}rez-Torres}, {P{\'e}rez-Gonz{\'a}lez}, {Huang}, {Davis},
  {Guhathakurta}, {Barmby}, {Conselice}, {Lozano}, {Newman}, \&
  {Cooper}}]{2011MNRAS.410.2556D}
{Dom{\'\i}nguez}, A., {Primack}, J.~R., {Rosario}, D.~J., {et~al.} 2011,
  \mnras, 410, 2556, \dodoi{10.1111/j.1365-2966.2010.17631.x}

\bibitem[{{Foffano} {et~al.}(2019){Foffano}, {Prandini}, {Franceschini}, \&
  {Paiano}}]{2019MNRAS.486.1741F}
{Foffano}, L., {Prandini}, E., {Franceschini}, A., \& {Paiano}, S. 2019,
  \mnras, 486, 1741, \dodoi{10.1093/mnras/stz812}

\bibitem[{{Francis} {et~al.}(1991){Francis}, {Hewett}, {Foltz}, {Chaffee},
  {Weymann}, \& {Morris}}]{1991ApJ...373..465F}
{Francis}, P.~J., {Hewett}, P.~C., {Foltz}, C.~B., {et~al.} 1991, \apj, 373,
  465, \dodoi{10.1086/170066}

\bibitem[{{Franckowiak} {et~al.}(2020){Franckowiak}, {Garrappa}, {Paliya},
  {Shappee}, {Stein}, {Strotjohann}, {Kowalski}, {Buson}, {Kiehlmann},
  {Max-Moerbeck}, \& {Angioni}}]{2020ApJ...893..162F}
{Franckowiak}, A., {Garrappa}, S., {Paliya}, V., {et~al.} 2020, \apj, 893, 162,
  \dodoi{10.3847/1538-4357/ab8307}

\bibitem[{{Garrappa} {et~al.}(2020){Garrappa}, {Buson}, \& {Fermi-LAT
  Collaboration}}]{2020GCN.26669....1G}
{Garrappa}, S., {Buson}, S., \& {Fermi-LAT Collaboration}. 2020, GRB
  Coordinates Network, 26669, 1

\bibitem[{{Garrappa} {et~al.}(2019){Garrappa}, {Buson}, {Franckowiak},
  {Fermi-LAT collaboration}, {Shappee}, {Beacom}, {Dong}, {Holoien},
  {Kochanek}, {Prieto}, {Stanek}, {Thompson}, {ASAS-SN collaboration},
  {Aartsen}, {Ackermann}, {Adams}, {Aguilar}, {Ahlers}, {Ahrens}, {Alispach},
  {Andeen}, {Anderson}, {Ansseau}, {Anton}, {Arg{\"u}elles}, {Auffenberg},
  {Axani}, {Backes}, {Bagherpour}, {Bai}, {Barbano}, {Barwick}, {Baum}, {Bay},
  {Beatty}, {Becker}, {Becker Tjus}, {BenZvi}, {Berley}, {Bernardini},
  {Besson}, {Binder}, {Bindig}, {Blaufuss}, {Blot}, {Bohm}, {B{\"o}rner},
  {B{\"o}ser}, {Botner}, {Bourbeau}, {Bourbeau}, {Bradascio}, {Braun}, {Bretz},
  {Bron}, {Brostean-Kaiser}, {Burgman}, {Busse}, {Carver}, {Chen}, {Cheung},
  {Chirkin}, {Clark}, {Classen}, {Collin}, {Conrad}, {Coppin}, {Correa},
  {Cowen}, {Cross}, {Dave}, {de Andr{\'e}}, {De Clercq}, {DeLaunay},
  {Dembinski}, {Deoskar}, {De Ridder}, {Desiati}, {de Vries}, {de Wasseige},
  {de With}, {DeYoung}, {Diaz}, {D{\'\i}az-V{\'e}lez}, {Dujmovic}, {Dunkman},
  {Dvorak}, {Eberhardt}, {Ehrhardt}, {Eller}, {Evenson}, {Fahey}, {Fazely},
  {Felde}, {Filimonov}, {Finley}, {Franckowiak}, {Friedman}, {Fritz},
  {Gaisser}, {Gallagher}, {Ganster}, {Garrappa}, {Gerhardt}, {Ghorbani},
  {Glauch}, {Gl{\"u}senkamp}, {Goldschmidt}, {Gonzalez}, {Grant}, {Griffith},
  {G{\"u}nder}, {G{\"u}nd{\"u}z}, {Haack}, {Hallgren}, {Halve}, {Halzen},
  {Hanson}, {Hebecker}, {Heereman}, {Helbing}, {Hellauer}, {Henningsen},
  {Hickford}, {Hignight}, {Hill}, {Hoffman}, {Hoffmann}, {Hoinka},
  {Hokanson-Fasig}, {Hoshina}, {Huang}, {Huber}, {Hultqvist}, {H{\"u}nnefeld},
  {Hussain}, {In}, {Iovine}, {Ishihara}, {Jacobi}, {Japaridze}, {Jeong},
  {Jero}, {Jones}, {Kang}, {Kappes}, {Kappesser}, {Karg}, {Karl}, {Karle},
  {Katz}, {Kauer}, {Keivani}, {Kelley}, {Kheirand ish}, {Kim}, {Kintscher},
  {Kiryluk}, {Kittler}, {Klein}, {Koirala}, {Kolanoski}, {K{\"o}pke}, {Kopper},
  {Kopper}, {Koskinen}, {Kowalski}, {Krings}, {Kr{\"u}ckl}, {Kulacz}, {Kunwar},
  {Kurahashi}, {Kyriacou}, {Labare}, {Lanfranchi}, {Larson}, {Lauber}, {Lazar},
  {Leonard}, {Leuermann}, {Liu}, {Lohfink}, {Lozano Mariscal}, {Lu},
  {Lucarelli}, {L{\"u}nemann}, {Luszczak}, {Madsen}, {Maggi}, {Mahn}, {Makino},
  {Mallot}, {Mancina}, {Mari{\textcommabelow s}}, {Maruyama}, {Mase}, {Maunu},
  {Meagher}, {Medici}, {Medina}, {Meier}, {Meighen-Berger}, {Menne}, {Merino},
  {Meures}, {Miarecki}, {Micallef}, {Moment{\'e}}, {Montaruli}, {Moore},
  {Moulai}, {Nagai}, {Nahnhauer}, {Nakarmi}, {Naumann}, {Neer}, {Niederhausen},
  {Nowicki}, {Nygren}, {Obertacke Pollmann}, {Olivas}, {O'Murchadha},
  {O'Sullivan}, {Palczewski}, {Pandya}, {Pankova}, {Park}, {Peiffer},
  {P{\'e}rez de los Heros}, {Pieloth}, {Pinat}, {Pizzuto}, {Plum}, {Price},
  {Przybylski}, {Raab}, {Raissi}, {Rameez}, {Rauch}, {Rawlins}, {Rea},
  {Reimann}, {Relethford}, {Renzi}, {Resconi}, {Rhode}, {Richman}, {Robertson},
  {Rongen}, {Rott}, {Ruhe}, {Ryckbosch}, {Rysewyk}, {Safa}, {Sanchez Herrera},
  {Sandrock}, {Sandroos}, {Santander}, {Sarkar}, {Sarkar}, {Satalecka},
  {Schaufel}, {Schlunder}, {Schmidt}, {Schneider}, {Schneider}, {Schumacher},
  {Sclafani}, {Seckel}, {Seunarine}, {Silva}, {Snihur}, {Soedingrekso},
  {Soldin}, {Song}, {Spiczak}, {Spiering}, {Stachurska}, {Stamatikos},
  {Stanev}, {Stasik}, {Stein}, {Stettner}, {Steuer}, {Stezelberger},
  {Stokstad}, {St{\"o}{\ss}l}, {Strotjohann}, {Stuttard}, {Sullivan},
  {Sutherland}, {Taboada}, {Tenholt}, {Ter-Antonyan}, {Terliuk}, {Tilav},
  {Tomankova}, {T{\"o}nnis}, {Toscano}, {Tosi}, {Tselengidou}, {Tung},
  {Turcati}, {Turcotte}, {Turley}, {Ty}, {Unger}, {Unland Elorrieta}, {Usner},
  {Vand enbroucke}, {Van Driessche}, {van Eijk}, {van Eijndhoven}, {Vanheule},
  {van Santen}, {Vraeghe}, {Walck}, {Wallace}, {Wallraff}, {Wandkowsky},
  {Watson}, {Weaver}, {Weiss}, {Weldert}, {Wendt}, {Werthebach}, {Westerhoff},
  {Whelan}, {Whitehorn}, {Wiebe}, {Wiebusch}, {Wille}, {Williams}, {Wills},
  {Wolf}, {Wood}, {Wood}, {Woschnagg}, {Wrede}, {Xu}, {Xu}, {Xu}, {Yanez},
  {Yodh}, {Yoshida}, {Yuan}, \& {IceCube Collaboration}}]{2019ApJ...880..103G}
{Garrappa}, S., {Buson}, S., {Franckowiak}, A., {et~al.} 2019, The
  Astrophysical Journal, 880, 103, \dodoi{10.3847/1538-4357/ab2ada}

\bibitem[{{Ghisellini} {et~al.}(2013){Ghisellini}, {Tavecchio}, {Foschini},
  {Bonnoli}, \& {Tagliaferri}}]{2013MNRAS.432L..66G}
{Ghisellini}, G., {Tavecchio}, F., {Foschini}, L., {Bonnoli}, G., \&
  {Tagliaferri}, G. 2013, \mnras, 432, L66, \dodoi{10.1093/mnrasl/slt041}

\bibitem[{{Giommi} {et~al.}(2020){Giommi}, {Glauch}, \&
  {Resconi}}]{2020ATel13394....1G}
{Giommi}, P., {Glauch}, T., \& {Resconi}, E. 2020, The Astronomer's Telegram,
  13394, 1

\bibitem[{{Giommi} {et~al.}(2012){Giommi}, {Polenta}, {L{\"a}hteenm{\"a}ki},
  {Thompson}, {Capalbi}, {Cutini}, {Gasparrini}, {Gonz{\'a}lez-Nuevo},
  {Le{\'o}n-Tavares}, {L{\'o}pez-Caniego}, {Mazziotta}, {Monte}, {Perri},
  {Rain{\`o}}, {Tosti}, {Tramacere}, {Verrecchia}, {Aller}, {Aller},
  {Angelakis}, {Bastieri}, {Berdyugin}, {Bonaldi}, {Bonavera}, {Burigana},
  {Burrows}, {Buson}, {Cavazzuti}, {Chincarini}, {Colafrancesco}, {Costamante},
  {Cuttaia}, {D'Ammando}, {de Zotti}, {Frailis}, {Fuhrmann}, {Galeotta},
  {Gargano}, {Gehrels}, {Giglietto}, {Giordano}, {Giroletti}, {Keih{\"a}nen},
  {King}, {Krichbaum}, {Lasenby}, {Lavonen}, {Lawrence}, {Leto}, {Lindfors},
  {Mandolesi}, {Massardi}, {Max-Moerbeck}, {Michelson}, {Mingaliev}, {Natoli},
  {Nestoras}, {Nieppola}, {Nilsson}, {Partridge}, {Pavlidou}, {Pearson},
  {Procopio}, {Rachen}, {Readhead}, {Reeves}, {Reimer}, {Reinthal},
  {Ricciardi}, {Richards}, {Riquelme}, {Saarinen}, {Sajina}, {Sandri},
  {Savolainen}, {Sievers}, {Sillanp{\"a}{\"a}}, {Sotnikova}, {Stevenson},
  {Tagliaferri}, {Takalo}, {Tammi}, {Tavagnacco}, {Terenzi}, {Toffolatti},
  {Tornikoski}, {Trigilio}, {Turunen}, {Umana}, {Ungerechts}, {Villa}, {Wu},
  {Zacchei}, {Zensus}, \& {Zhou}}]{2012A&A...541A.160G}
{Giommi}, P., {Polenta}, G., {L{\"a}hteenm{\"a}ki}, A., {et~al.} 2012, \aap,
  541, A160, \dodoi{10.1051/0004-6361/201117825}

\bibitem[{{G{\"u}ltekin} {et~al.}(2009){G{\"u}ltekin}, {Richstone}, {Gebhardt},
  {Lauer}, {Tremaine}, {Aller}, {Bender}, {Dressler}, {Faber}, {Filippenko},
  {Green}, {Ho}, {Kormendy}, {Magorrian}, {Pinkney}, \&
  {Siopis}}]{2009ApJ...698..198G}
{G{\"u}ltekin}, K., {Richstone}, D.~O., {Gebhardt}, K., {et~al.} 2009, \apj,
  698, 198, \dodoi{10.1088/0004-637X/698/1/198}

\bibitem[{{IceCube Collaboration}(2020)}]{2020GCN.26655....1I}
{IceCube Collaboration}. 2020, GRB Coordinates Network, 26655, 1

\bibitem[{{IceCube Collaboration} {et~al.}(2018{\natexlab{a}}){IceCube
  Collaboration}, {Aartsen}, {Ackermann}, {Adams}, {Aguilar}, {Ahlers},
  {Ahrens}, {Al Samarai}, {Altmann}, {Andeen}, {Anderson}, {Ansseau}, {Anton},
  {Arg{\"u}elles}, {Auffenberg}, {Axani}, {Bagherpour}, {Bai}, {Barron},
  {Barwick}, {Baum}, {Bay}, {Beatty}, {Becker Tjus}, {Becker}, {BenZvi},
  {Berley}, {Bernardini}, {Besson}, {Binder}, {Bindig}, {Blaufuss}, {Blot},
  {Bohm}, {B{\"o}rner}, {Bos}, {B{\"o}ser}, {Botner}, {Bourbeau}, {Bourbeau},
  {Bradascio}, {Braun}, {Brenzke}, {Bretz}, {Bron}, {Brostean-Kaiser},
  {Burgman}, {Busse}, {Carver}, {Cheung}, {Chirkin}, {Christov}, {Clark},
  {Classen}, {Coenders}, {Collin}, {Conrad}, {Coppin}, {Correa}, {Cowen},
  {Cross}, {Dave}, {Day}, {de Andr{\'e}}, {De Clercq}, {DeLaunay}, {Dembinski},
  {De Ridder}, {Desiati}, {de Vries}, {de Wasseige}, {de With}, {DeYoung},
  {D{\'\i}az-V{\'e}lez}, {di Lorenzo}, {Dujmovic}, {Dumm}, {Dunkman}, {Dvorak},
  {Eberhardt}, {Ehrhardt}, {Eichmann}, {Eller}, {Evenson}, {Fahey}, {Fazely},
  {Felde}, {Filimonov}, {Finley}, {Flis}, {Franckowiak}, {Friedman}, {Fritz},
  {Gaisser}, {Gallagher}, {Gerhardt}, {Ghorbani}, {Glauch}, {Gl{\"u}senkamp},
  {Goldschmidt}, {Gonzalez}, {Grant}, {Griffith}, {Haack}, {Hallgren},
  {Halzen}, {Hanson}, {Hebecker}, {Heereman}, {Helbing}, {Hellauer},
  {Hickford}, {Hignight}, {Hill}, {Hoffman}, {Hoffmann}, {Hoinka},
  {Hokanson-Fasig}, {Hoshina}, {Huang}, {Huber}, {Hultqvist}, {H{\"u}nnefeld},
  {Hussain}, {In}, {Iovine}, {Ishihara}, {Jacobi}, {Japaridze}, {Jeong},
  {Jero}, {Jones}, {Kalaczynski}, {Kang}, {Kappes}, {Kappesser}, {Karg},
  {Karle}, {Katz}, {Kauer}, {Keivani}, {Kelley}, {Kheirandish}, {Kim}, {Kim},
  {Kintscher}, {Kiryluk}, {Kittler}, {Klein}, {Koirala}, {Kolanoski},
  {K{\"o}pke}, {Kopper}, {Kopper}, {Koschinsky}, {Koskinen}, {Kowalski},
  {Krings}, {Kroll}, {Kr{\"u}ckl}, {Kunwar}, {Kurahashi}, {Kuwabara},
  {Kyriacou}, {Labare}, {Lanfranchi}, {Larson}, {Lauber}, {Leonard},
  {Lesiak-Bzdak}, {Leuermann}, {Liu}, {Lozano Mariscal}, {Lu}, {L{\"u}nemann},
  {Luszczak}, {Madsen}, {Maggi}, {Mahn}, {Mancina}, {Maruyama}, {Mase},
  {Maunu}, {Meagher}, {Medici}, {Meier}, {Menne}, {Merino}, {Meures},
  {Miarecki}, {Micallef}, {Moment{\'e}}, {Montaruli}, {Moore}, {S}, {Morse},
  {Moulai}, {Nahnhauer}, {Nakarmi}, {Naumann}, {Neer}, {Niederhausen},
  {Nowicki}, {Nygren}, {Obertacke Pollmann}, {Olivas}, {O'Murchadha},
  {O'Sullivan}, {Palczewski}, {Pandya}, {Pankova}, {Peiffer}, {Pepper},
  {P{\'e}rez de los Heros}, {Pieloth}, {Pinat}, {Plum}, {Price}, {Przybylski},
  {Raab}, {R{\"a}del}, {Rameez}, {Rauch}, {Rawlins}, {Rea}, {Reimann},
  {Relethford}, {Relich}, {Resconi}, {Rhode}, {Richman}, {Robertson}, {Rongen},
  {Rott}, {Ruhe}, {Ryckbosch}, {Rysewyk}, {Safa}, {S{\"a}lzer}, {Sanchez
  Herrera}, {Sandrock}, {Sandroos}, {Santander}, {Sarkar}, {Sarkar},
  {Satalecka}, {Schlunder}, {Schmidt}, {Schneider}, {Schoenen},
  {Sch{\"o}neberg}, {Schumacher}, {Sclafani}, {Seckel}, {Seunarine},
  {Soedingrekso}, {Soldin}, {Song}, {Spiczak}, {Spiering}, {Stachurska},
  {Stamatikos}, {Stanev}, {Stasik}, {Stein}, {Stettner}, {Steuer},
  {Stezelberger}, {Stokstad}, {St{\"o}{\ss}l}, {Strotjohann}, {Stuttard},
  {Sullivan}, {Sutherland}, {Taboada}, {Tatar}, {Tenholt}, {Ter-Antonyan},
  {Terliuk}, {Tilav}, {Toale}, {Tobin}, {Toennis}, {Toscano}, {Tosi},
  {Tselengidou}, {Tung}, {Turcati}, {Turley}, {Ty}, {Unger}, {Usner},
  {Vandenbroucke}, {Van Driessche}, {van Eijk}, {van Eijndhoven}, {Vanheule},
  {van Santen}, {Vogel}, {Vraeghe}, {Walck}, {Wallace}, {Wallraff}, {Wandler},
  {Wandkowsky}, {Waza}, {Weaver}, {Weiss}, {Wendt}, {Werthebach}, {Westerhoff},
  {Whelan}, {Whitehorn}, {Wiebe}, {Wiebusch}, {Wille}, {Williams}, {Wills},
  {Wolf}, {Wood}, {Wood}, {Woschnagg}, {Xu}, {Xu}, {Xu}, {Yanez}, {Yodh},
  {Yoshida}, {Yuan}, {Fermi-LAT Collaboration}, {Abdollahi}, {Ajello},
  {Angioni}, {Baldini}, {Ballet}, {Barbiellini}, {Bastieri}, {Bechtol},
  {Bellazzini}, {Berenji}, {Bissaldi}, {Blandford}, {Bonino}, {Bottacini},
  {Bregeon}, {Bruel}, {Buehler}, {Burnett}, {Burns}, {Buson}, {Cameron},
  {Caputo}, {Caraveo}, {Cavazzuti}, {Charles}, {Chen}, {Cheung}, {Chiang},
  {Chiaro}, {Ciprini}, {Cohen-Tanugi}, {Conrad}, {Costantin}, {Cutini},
  {D'Ammando}, {de Palma}, {Digel}, {Di Lalla}, {Di Mauro}, {Di Venere},
  {Dom{\'\i}nguez}, {Favuzzi}, {Franckowiak}, {Fukazawa}, {Funk}, {Fusco},
  {Gargano}, {Gasparrini}, {Giglietto}, {Giomi}, {Giommi}, {Giordano},
  {Giroletti}, {Glanzman}, {Green}, {Grenier}, {Grondin}, {Guiriec}, {Harding},
  {Hayashida}, {Hays}, {Hewitt}, {Horan}, {J{\'o}hannesson}, {Kadler},
  {Kensei}, {Kocevski}, {Krauss}, {Kreter}, {Kuss}, {La Mura}, {Larsson},
  {Latronico}, {Lemoine-Goumard}, {Li}, {Longo}, {Loparco}, {Lovellette},
  {Lubrano}, {Magill}, {Maldera}, {Malyshev}, {Manfreda}, {Mazziotta},
  {McEnery}, {Meyer}, {Michelson}, {Mizuno}, {Monzani}, {Morselli},
  {Moskalenko}, {Negro}, {Nuss}, {Ojha}, {Omodei}, {Orienti}, {Orlando},
  {Palatiello}, {Paliya}, {Perkins}, {Persic}, {Pesce-Rollins}, {Piron},
  {Porter}, {Principe}, {Rain{\`o}}, {Rando}, {Rani}, {Razzano}, {Razzaque},
  {Reimer}, {Reimer}, {Renault-Tinacci}, {Ritz}, {Rochester}, {Saz Parkinson},
  {Sgr{\`o}}, {Siskind}, {Spandre}, {Spinelli}, {Suson}, {Tajima}, {Takahashi},
  {Tanaka}, {Thayer}, {Thompson}, {Tibaldo}, {Torres}, {Torresi}, {Tosti},
  {Troja}, {Valverde}, {Vianello}, {Vogel}, {Wood}, {Wood}, {Zaharijas}, {MAGIC
  Collaboration}, {Ahnen}, {Ansoldi}, {Antonelli}, {Arcaro}, {Baack},
  {Babi{\'c}}, {Banerjee}, {Bangale}, {Barres de Almeida}, {Barrio}, {Becerra
  Gonz{\'a}lez}, {Bednarek}, {Bernardini}, {Berti}, {Bhattacharyya}, {Biland},
  {Blanch}, {Bonnoli}, {Carosi}, {Carosi}, {Ceribella}, {Chatterjee}, {Colak},
  {Colin}, {Colombo}, {Contreras}, {Cortina}, {Covino}, {Cumani}, {Da Vela},
  {Dazzi}, {De Angelis}, {De Lotto}, {Delfino}, {Delgado}, {Di Pierro},
  {Dom{\'\i}nguez}, {Dominis Prester}, {Dorner}, {Doro}, {Einecke},
  {Elsaesser}, {Fallah Ramazani}, {Fern{\'a}ndez-Barral}, {Fidalgo}, {Foffano},
  {Pfrang}, {Fonseca}, {Font}, {Franceschini}, {Fruck}, {Galindo}, {Gallozzi},
  {Garc{\'\i}a L{\'o}pez}, {Garczarczyk}, {Gaug}, {Giammaria}, {Godinovi{\'c}},
  {Gora}, {Guberman}, {Hadasch}, {Hahn}, {Hassan}, {Hayashida}, {Herrera},
  {Hose}, {Hrupec}, {Inoue}, {Ishio}, {Konno}, {Kubo}, {Kushida}, {Lelas},
  {Lindfors}, {Lombardi}, {Longo}, {L{\'o}pez}, {Maggio}, {Majumdar},
  {Makariev}, {Maneva}, {Manganaro}, {Mannheim}, {Maraschi}, {Mariotti},
  {Mart{\'\i}nez}, {Masuda}, {Mazin}, {Minev}, {M}, {Mirzoyan}, {Moralejo},
  {Moreno}, {Moretti}, {Nagayoshi}, {Neustroev}, {Niedzwiecki}, {Nievas
  Rosillo}, {Nigro}, {Nilsson}, {Ninci}, {Nishijima}, {Noda}, {Nogu{\'e}s},
  {Paiano}, {Palacio}, {Paneque}, {Paoletti}, {Paredes}, {Pedaletti},
  {Peresano}, {Persic}, {Prada Moroni}, {Prandini}, {Puljak}, {Rodriguez
  Garcia}, {Reichardt}, {Rhode}, {Rib{\'o}}, {Rico}, {Righi}, {Rugliancich},
  {Saito}, {Satalecka}, {Schweizer}, {Sitarek}, {{\v{S}}nidaric ́},
  {Sobczynska}, {Stamerra}, {Strzys}, {Suri{\'c}}, {Takahashi}, {Tavecchio},
  {Temnikov}, {Terzi{\'c}}, {Teshima}, {Torres-Alb{\`a}}, {Treves},
  {Tsujimoto}, {Vanzo}, {Vazquez Acosta}, {Vovk}, {Ward}, {Will}, {S}, {Zaric
  ́}, {AGILE Team}, {Lucarelli}, {Tavani}, {Piano}, {Donnarumma}, {Pittori},
  {Verrecchia}, {Barbiellini}, {Bulgarelli}, {Caraveo}, {Cattaneo},
  {Colafrancesco}, {Costa}, {Di Cocco}, {Ferrari}, {Gianotti}, {Giuliani},
  {Lipari}, {Mereghetti}, {Morselli}, {Pacciani}, {Paoletti}, {Parmiggiani},
  {Pellizzoni}, {Picozza}, {Pilia}, {Rappoldi}, {Trois}, {Vercellone},
  {Vittorini}, {ASAS-SN Team}, {Stanek}, {Kochanek}, {Beacom}, {Thompson},
  {Holoien}, {Dong}, {Prieto}, {Shappee}, {Holmbo}, {HAWC Collaboration},
  {Abeysekara}, {Albert}, {Alfaro}, {Alvarez}, {Arceo},
  {Arteaga-Vel{\'a}zquez}, {Avila Rojas}, {Ayala Solares}, {Becerril},
  {Belmont-Moreno}, {Bernal}, {Caballero-Mora}, {Capistr{\'a}n},
  {Carrami{\~n}ana}, {Casanova}, {Castillo}, {Cotti}, {Cotzomi}, {Couti{\~n}o
  de Le{\'o}n}, {De Le{\'o}n}, {De la Fuente}, {Diaz Hernandez}, {Dichiara},
  {Dingus}, {DuVernois}, {D{\'\i}az-V{\'e}lez}, {Ellsworth}, {Engel},
  {Fiorino}, {Fleischhack}, {Fraija}, {Garc{\'\i}a-Gonz{\'a}lez}, {Garfias},
  {Gonz{\'a}lez Mu{\~n}oz}, {Gonz{\'a}lez}, {Goodman}, {Hampel-Arias},
  {Harding}, {Hernand ez}, {Hona}, {Hueyotl-Zahuantitla}, {Hui},
  {H{\"u}ntemeyer}, {Iriarte}, {Jardin-Blicq}, {Joshi}, {Kaufmann}, {Kunde},
  {Lara}, {Lauer}, {Lee}, {Lennarz}, {Le{\'o}n Vargas}, {Linnemann},
  {Longinotti}, {Luis-Raya}, {Luna-Garc{\'\i}a}, {Malone}, {Marinelli},
  {Martinez}, {Martinez-Castellanos}, {Mart{\'\i}nez-Castro},
  {Mart{\'\i}nez-Huerta}, {Matthews}, {Miranda-Romagnoli}, {Moreno},
  {Mostaf{\'a}}, {Nayerhoda}, {Nellen}, {Newbold}, {Nisa}, {Noriega-Papaqui},
  {Pelayo}, {Pretz}, {P{\'e}rez-P{\'e}rez}, {Ren}, {Rho}, {Rivi{\`e}re},
  {Rosa-Gonz{\'a}lez}, {Rosenberg}, {Ruiz-Velasco}, {Ruiz-Velasco}, {Salesa
  Greus}, {Sandoval}, {Schneider}, {Schoorlemmer}, {Sinnis}, {Smith},
  {Springer}, {Surajbali}, {Tibolla}, {Tollefson}, {Torres}, {Villase{\~n}or},
  {Weisgarber}, {Werner}, {Yapici}, {Gaurang}, {Zepeda}, {Zhou}, {{\'A}lvarez},
  {H.~E.~S.~S. Collaboration}, {Abdalla}, {Ang{\"u}ner}, {Armand}, {Backes},
  {Becherini}, {Berge}, {B{\"o}ttcher}, {Boisson}, {Bolmont}, {Bonnefoy},
  {Bordas}, {Brun}, {B{\"u}chele}, {Bulik}, {Caroff}, {Carosi}, {Casanova},
  {Cerruti}, {Chakraborty}, {Chandra}, {Chen}, {Colafrancesco}, {Davids},
  {Deil}, {Devin}, {Djannati-Ata{\"\i}}, {Egberts}, {Emery}, {Eschbach},
  {Fiasson}, {Fontaine}, {Funk}, {F{\"u}{\ss}ling}, {Gallant}, {Gat{\'e}},
  {Giavitto}, {Glawion}, {Glicenstein}, {Gottschall}, {Grondin}, {Haupt},
  {Henri}, {Hinton}, {Hoischen}, {Holch}, {Huber}, {Jamrozy}, {Jankowsky},
  {Jankowsky}, {Jouvin}, {Jung-Richardt}, {Kerszberg}, {Kh{\'e}lifi}, {King},
  {Klepser}, {Kluz ́niak}, {Komin}, {Kraus}, {Lefaucheur}, {Lemi{\`e}re},
  {Lemoine-Goumard}, {Lenain}, {Leser}, {Lohse}, {L{\'o}pez-Coto}, {Lorentz},
  {Lypova}, {Marandon}, {Guillem Mart{\'\i}-Devesa}, {Maurin}, {Mitchell},
  {Moderski}, {Mohamed}, {Mohrmann}, {Moulin}, {Murach}, {de Naurois},
  {Niederwanger}, {Niemiec}, {Oakes}, {O'Brien}, {Ohm}, {Ostrowski}, {Oya},
  {Panter}, {Parsons}, {Perennes}, {Piel}, {Pita}, {Poireau}, {Priyana Noel},
  {Prokoph}, {P{\"u}hlhofer}, {Quirrenbach}, {Raab}, {Rauth}, {Renaud},
  {Rieger}, {Rinchiuso}, {Romoli}, {Rowell}, {Rudak}, {Sasaki}, {Sanchez},
  {Schlickeiser}, {Sch{\"u}ssler}, {Schulz}, {Schwanke}, {Seglar-Arroyo},
  {Shafi}, {Simoni}, {Sol}, {Stegmann}, {Steppa}, {Tavernier}, {Taylor},
  {Tiziani}, {Trichard}, {Tsirou}, {van Eldik}, {van Rensburg}, {van Soelen},
  {Veh}, {Vincent}, {Voisin}, {Wagner}, {Wagner}, {Wierzcholska}, {Zanin},
  {Zdziarski}, {Zech}, {Ziegler}, {Zorn}, {{\.Z}ywucka}, {INTEGRAL Team},
  {Savchenko}, {Ferrigno}, {Bazzano}, {Diehl}, {Kuulkers}, {Laurent},
  {Mereghetti}, {Natalucci}, {Panessa}, {Rodi}, {Ubertini}, {Kanata}, Teams,
  {Morokuma}, {Ohta}, {Tanaka}, {Mori}, {Yamanaka}, {Kawabata}, {Utsumi},
  {Nakaoka}, {Kawabata}, {Nagashima}, {Yoshida}, {Matsuoka}, {Itoh}, {Kapteyn
  Team}, {Keel}, {Liverpool Telescope Team}, {Copperwheat}, {Steele},
  {Swift/NuSTAR Team}, {Cenko}, {Cowen}, {DeLaunay}, {Evans}, {Fox}, {Keivani},
  {Kennea}, {Marshall}, {Osborne}, {Santander}, {Tohuvavohu}, {Turley},
  {VERITAS Collaboration}, {Abeysekara}, {Archer}, {Benbow}, {Bird}, {Brill},
  {Brose}, {Buchovecky}, {Buckley}, {Bugaev}, {Christiansen}, {Connolly},
  {Cui}, {Daniel}, {Errando}, {Falcone}, {Feng}, {Finley}, {Fortson},
  {Furniss}, {Gueta}, {H{\"u}tten}, {Hervet}, {Hughes}, {Humensky}, {Johnson},
  {Kaaret}, {Kar}, {Kelley-Hoskins}, {Kertzman}, {Kieda}, {Krause},
  {Krennrich}, {Kumar}, {Lang}, {Lin}, {Maier}, {McArthur}, {Moriarty},
  {Mukherjee}, {Nieto}, {O'Brien}, {Ong}, {Otte}, {Park}, {Petrashyk}, {Pohl},
  {Popkow}, {Pueschel}, {Quinn}, {Ragan}, {Reynolds}, {Richards}, {Roache},
  {Rulten}, {Sadeh}, {Santander}, {Scott}, {Sembroski}, {Shahinyan}, {Sushch},
  {Tr{\'e}panier}, {Tyler}, {Vassiliev}, {Wakely}, {Weinstein}, {Wells},
  {Wilcox}, {Wilhelm}, {Williams}, {Zitzer}, {VLA/B Team}, {Tetarenko},
  {Kimball}, {Miller-Jones}, \& {Sivakoff}}]{2018Sci...361.1378I}
{IceCube Collaboration}, {Aartsen}, M.~G., {Ackermann}, M., {et~al.}
  2018{\natexlab{a}}, Science, 361, eaat1378, \dodoi{10.1126/science.aat1378}

\bibitem[{{IceCube Collaboration} {et~al.}(2018{\natexlab{b}}){IceCube
  Collaboration}, {Aartsen}, {Ackermann}, {Adams}, {Aguilar}, {Ahlers},
  {Ahrens}, {Samarai}, {Altmann}, {Andeen}, {Anderson}, {Ansseau}, {Anton},
  {Arg{\"u}elles}, {Arsioli}, {Auffenberg}, {Axani}, {Bagherpour}, {Bai},
  {Barron}, {Barwick}, {Baum}, {Bay}, {Beatty}, {Becker Tjus}, {Becker},
  {BenZvi}, {Berley}, {Bernardini}, {Besson}, {Binder}, {Bindig}, {Blaufuss},
  {Blot}, {Bohm}, {B{\"o}rner}, {Bos}, {B{\"o}ser}, {Botner}, {Bourbeau},
  {Bourbeau}, {Bradascio}, {Braun}, {Brenzke}, {Bretz}, {Bron},
  {Brostean-Kaiser}, {Burgman}, {Busse}, {Carver}, {Cheung}, {Chirkin},
  {Christov}, {Clark}, {Classen}, {Coenders}, {Collin}, {Conrad}, {Coppin},
  {Correa}, {Cowen}, {Cross}, {Dave}, {Day}, {de Andr{\'e}}, {De Clercq},
  {DeLaunay}, {Dembinski}, {DeRidder}, {Desiati}, {de Vries}, {de Wasseige},
  {de With}, {DeYoung}, {D{\'\i}az-V{\'e}lez}, {di Lorenzo}, {Dujmovic},
  {Dumm}, {Dunkman}, {Dvorak}, {Eberhardt}, {Ehrhardt}, {Eichmann}, {Eller},
  {Evenson}, {Fahey}, {Fazely}, {Felde}, {Filimonov}, {Finley}, {Flis},
  {Franckowiak}, {Friedman}, {Fritz}, {Gaisser}, {Gallagher}, {Gerhardt},
  {Ghorbani}, {Giommi}, {Glauch}, {Gl{\"u}senkamp}, {Goldschmidt}, {Gonzalez},
  {Grant}, {Griffith}, {Haack}, {Hallgren}, {Halzen}, {Hanson}, {Hebecker},
  {Heereman}, {Helbing}, {Hellauer}, {Hickford}, {Hignight}, {Hill}, {Hoffman},
  {Hoffmann}, {Hoinka}, {Hokanson-Fasig}, {Hoshina}, {Huang}, {Huber},
  {Hultqvist}, {H{\"u}nnefeld}, {Hussain}, {In}, {Iovine}, {Ishihara},
  {Jacobi}, {Japaridze}, {Jeong}, {Jero}, {Jones}, {Kalaczynski}, {Kang},
  {Kappes}, {Kappesser}, {Karg}, {Karle}, {Katz}, {Kauer}, {Keivani}, {Kelley},
  {Kheirandish}, {Kim}, {Kim}, {Kintscher}, {Kiryluk}, {Kittler}, {Klein},
  {Koirala}, {Kolanoski}, {K{\"o}pke}, {Kopper}, {Kopper}, {Koschinsky},
  {Koskinen}, {Kowalski}, {Krammer}, {Krings}, {Kroll}, {Kr{\"u}ckl}, {Kunwar},
  {Kurahashi}, {Kuwabara}, {Kyriacou}, {Labare}, {Lanfranchi}, {Larson},
  {Lauber}, {Leonard}, {Lesiak-Bzdak}, {Leuermann}, {Liu}, {Lozano Mariscal},
  {Lu}, {L{\"u}nemann}, {Luszczak}, {Madsen}, {Maggi}, {Mahn}, {Mancina},
  {Maruyama}, {Mase}, {Maunu}, {Meagher}, {Medici}, {Meier}, {Menne}, {Merino},
  {Meures}, {Miarecki}, {Micallef}, {Moment{\'e}}, {Montaruli}, {Moore},
  {Morse}, {Moulai}, {Nahnhauer}, {Nakarmi}, {Naumann}, {Neer}, {Niederhausen},
  {Nowicki}, {Nygren}, {Obertacke Pollmann}, {Olivas}, {O'Murchadha},
  {O'Sullivan}, {Padovani}, {Palczewski}, {Pand ya}, {Pankova}, {Peiffer},
  {Pepper}, {P{\'e}rez de los Heros}, {Pieloth}, {Pinat}, {Plum}, {Price},
  {Przybylski}, {Raab}, {R{\"a}del}, {Rameez}, {Rawlins}, {Rea}, {Reimann},
  {Relethford}, {Relich}, {Resconi}, {Rhode}, {Richman}, {Robertson}, {Rongen},
  {Rott}, {Ruhe}, {Ryckbosch}, {Rysewyk}, {Safa}, {Sahakyan}, {S{\"a}lzer},
  {Sanchez Herrera}, {Sandrock}, {Sandroos}, {Santander}, {Sarkar}, {Sarkar},
  {Satalecka}, {Schlunder}, {Schmidt}, {Schneider}, {Schoenen},
  {Sch{\"o}neberg}, {Schumacher}, {Sclafani}, {Seckel}, {Seunarine},
  {Soedingrekso}, {Soldin}, {Song}, {Spiczak}, {Spiering}, {Stachurska},
  {Stamatikos}, {Stanev}, {Stasik}, {Stettner}, {Steuer}, {Stezelberger},
  {Stokstad}, {St{\"o}{\ss}l}, {Strotjohann}, {Stuttard}, {Sullivan},
  {Sutherland}, {Taboada}, {Tatar}, {Tenholt}, {Ter-Antonyan}, {Terliuk},
  {Tilav}, {Toale}, {Tobin}, {Toennis}, {Toscano}, {Tosi}, {Tselengidou},
  {Tung}, {Turcati}, {Turley}, {Ty}, {Unger}, {Usner}, {Vandenbroucke}, {Van
  Driessche}, {van Eijk}, {van Eijndhoven}, {Vanheule}, {van Santen}, {Vogel},
  {Vraeghe}, {Walck}, {Wallace}, {Wallraff}, {Wandler}, {Wand kowsky}, {Waza},
  {Weaver}, {Weiss}, {Wendt}, {Werthebach}, {Westerhoff}, {Whelan},
  {Whitehorn}, {Wiebe}, {Wiebusch}, {Wille}, {Williams}, {Wills}, {Wolf},
  {Wood}, {Wood}, {Woschnagg}, {Xu}, {Xu}, {Xu}, {Yanez}, {Yodh}, {Yoshida}, \&
  {Yuan}}]{2018Sci...361..147I}
---. 2018{\natexlab{b}}, Science, 361, 147, \dodoi{10.1126/science.aat2890}

\bibitem[{{Kalberla} {et~al.}(2005){Kalberla}, {Burton}, {Hartmann}, {Arnal},
  {Bajaja}, {Morras}, \& {P{\"o}ppel}}]{Kalberla05}
{Kalberla}, P.~M.~W., {Burton}, W.~B., {Hartmann}, D., {et~al.} 2005, \aap,
  440, 775, \dodoi{10.1051/0004-6361:20041864}

\bibitem[{{Keivani} {et~al.}(2018){Keivani}, {Murase}, {Petropoulou}, {Fox},
  {Cenko}, {Chaty}, {Coleiro}, {DeLaunay}, {Dimitrakoudis}, {Evans}, {Kennea},
  {Marshall}, {Mastichiadis}, {Osborne}, {Santand er}, {Tohuvavohu}, \&
  {Turley}}]{2018ApJ...864...84K}
{Keivani}, A., {Murase}, K., {Petropoulou}, M., {et~al.} 2018, \apj, 864, 84,
  \dodoi{10.3847/1538-4357/aad59a}

\bibitem[{{Kelner} \& {Aharonian}(2008)}]{KA08}
{Kelner}, S.~R., \& {Aharonian}, F.~A. 2008, \prd, 78, 034013,
  \dodoi{10.1103/PhysRevD.78.034013}

\bibitem[{King(1985)}]{extinction}
King, D.~L. 1985, ING Technical Note, 31,
  \url{https://www.ing.iac.es/Astronomy/observing/manuals/ps/tech\_notes/tn031.pdf}

\bibitem[{{Kormendy} \& {Ho}(2013)}]{2013ARA&A..51..511K}
{Kormendy}, J., \& {Ho}, L.~C. 2013, \araa, 51, 511,
  \dodoi{10.1146/annurev-astro-082708-101811}

\bibitem[{{Krauss} {et~al.}(2020){Krauss}, {Gregoire}, {Fox}, {Kennea}, \&
  {Evans}}]{2020ATel13395....1K}
{Krauss}, F., {Gregoire}, T., {Fox}, D.~B., {Kennea}, J., \& {Evans}, P. 2020,
  The Astronomer's Telegram, 13395, 1

\bibitem[{{Kronmueller} \& {Glauch}(2019)}]{2019ICRC...36..937K}
{Kronmueller}, M., \& {Glauch}, T. 2019, in International Cosmic Ray
  Conference, Vol.~36, 36th International Cosmic Ray Conference (ICRC2019),
  937.
\newblock \doarXiv{1908.08763}

\bibitem[{{Lucarelli} {et~al.}(2019){Lucarelli}, {Tavani}, {Piano},
  {Bulgarelli}, {Donnarumma}, {Verrecchia}, {Pittori}, {Antonelli}, {Argan},
  {Barbiellini}, {Caraveo}, {Cardillo}, {Cattaneo}, {Chen}, {Colafrancesco},
  {Costa}, {Del Monte}, {Di Cocco}, {Ferrari}, {Fioretti}, {Galli}, {Giommi},
  {Giuliani}, {Lipari}, {Longo}, {Mereghetti}, {Morselli}, {Paoletti},
  {Parmiggiani}, {Pellizzoni}, {Picozza}, {Pilia}, {Rappoldi}, {Trois}, {Ursi},
  {Vercellone}, {Vittorini}, \& {AGILE Team}}]{2019ApJ...870..136L}
{Lucarelli}, F., {Tavani}, M., {Piano}, G., {et~al.} 2019, \apj, 870, 136,
  \dodoi{10.3847/1538-4357/aaf1c0}

\bibitem[{{Mannheim} {et~al.}(1992){Mannheim}, {Stanev}, \&
  {Biermann}}]{1992A&A...260L...1M}
{Mannheim}, K., {Stanev}, T., \& {Biermann}, P.~L. 1992, \aap, 260, L1

\bibitem[{{Merritt}(1997)}]{1997AJ....114..228M}
{Merritt}, D. 1997, \aj, 114, 228, \dodoi{10.1086/118467}

\bibitem[{{M{\"u}cke} {et~al.}(1999){M{\"u}cke}, {Rachen}, {Engel},
  {Protheroe}, \& {Stanev}}]{Muecke99}
{M{\"u}cke}, A., {Rachen}, J.~P., {Engel}, R., {Protheroe}, R.~J., \& {Stanev},
  T. 1999, \pasa, 16, 160, \dodoi{10.1071/AS99160}

\bibitem[{{Murase}(2017)}]{2017nacs.book...15M}
{Murase}, K. 2017, {Active Galactic Nuclei as High-Energy Neutrino Sources},
  ed. T.~{Gaisser} \& A.~{Karle}, 15--31, \dodoi{10.1142/9789814759410_0002}

\bibitem[{{Padovani} {et~al.}(2016){Padovani}, {Resconi}, {Giommi}, {Arsioli},
  \& {Chang}}]{2016MNRAS.457.3582P}
{Padovani}, P., {Resconi}, E., {Giommi}, P., {Arsioli}, B., \& {Chang}, Y.~L.
  2016, \mnras, 457, 3582, \dodoi{10.1093/mnras/stw228}

\bibitem[{{Paliya} {et~al.}(2019{\natexlab{a}}){Paliya}, {Dom{\'\i}nguez},
  {Ajello}, {Franckowiak}, \& {Hartmann}}]{2019ApJ...882L...3P}
{Paliya}, V.~S., {Dom{\'\i}nguez}, A., {Ajello}, M., {Franckowiak}, A., \&
  {Hartmann}, D. 2019{\natexlab{a}}, \apjl, 882, L3,
  \dodoi{10.3847/2041-8213/ab398a}

\bibitem[{{Paliya} {et~al.}(2017){Paliya}, {Marcotulli}, {Ajello}, {Joshi},
  {Sahayanathan}, {Rao}, \& {Hartmann}}]{2017ApJ...851...33P}
{Paliya}, V.~S., {Marcotulli}, L., {Ajello}, M., {et~al.} 2017, \apj, 851, 33,
  \dodoi{10.3847/1538-4357/aa98e1}

\bibitem[{{Paliya} {et~al.}(2019{\natexlab{b}}){Paliya}, {Koss},
  {Trakhtenbrot}, {Ricci}, {Oh}, {Ajello}, {Stern}, {Powell}, {Urry},
  {Harrison}, {Lamperti}, {Mushotzky}, {Marcotulli}, {Mej{\'\i}a-Restrepo}, \&
  {Hartmann}}]{2019ApJ...881..154P}
{Paliya}, V.~S., {Koss}, M., {Trakhtenbrot}, B., {et~al.} 2019{\natexlab{b}},
  \apj, 881, 154, \dodoi{10.3847/1538-4357/ab2f8b}

\bibitem[{{Petropoulou} {et~al.}(2015){Petropoulou}, {Dimitrakoudis},
  {Padovani}, {Mastichiadis}, \& {Resconi}}]{2015MNRAS.448.2412P}
{Petropoulou}, M., {Dimitrakoudis}, S., {Padovani}, P., {Mastichiadis}, A., \&
  {Resconi}, E. 2015, \mnras, 448, 2412, \dodoi{10.1093/mnras/stv179}

\bibitem[{{Petropoulou} {et~al.}(2020){Petropoulou}, {Oikonomou},
  {Mastichiadis}, {Murase}, {Padovani}, {Vasilopoulos}, \&
  {Giommi}}]{2020arXiv200507218P}
{Petropoulou}, M., {Oikonomou}, F., {Mastichiadis}, A., {et~al.} 2020, arXiv
  e-prints, arXiv:2005.07218.
\newblock \doarXiv{2005.07218}

\bibitem[{{Planck Collaboration} {et~al.}(2016){Planck Collaboration}, {Ade},
  {Aghanim}, {Arnaud}, {Ashdown}, {Aumont}, {Baccigalupi}, {Banday},
  {Barreiro}, {Bartlett}, \& et~al.}]{2016A&A...594A..13P}
{Planck Collaboration}, {Ade}, P.~A.~R., {Aghanim}, N., {et~al.} 2016, \aap,
  594, A13, \dodoi{10.1051/0004-6361/201525830}

\bibitem[{{Reimer} {et~al.}(2019){Reimer}, {B{\"o}ttcher}, \&
  {Buson}}]{Reimer19}
{Reimer}, A., {B{\"o}ttcher}, M., \& {Buson}, S. 2019, \apj, 881, 46,
  \dodoi{10.3847/1538-4357/ab2bff}

\bibitem[{{Rodrigues} {et~al.}(2019){Rodrigues}, {Gao}, {Fedynitch},
  {Palladino}, \& {Winter}}]{2019ApJ...874L..29R}
{Rodrigues}, X., {Gao}, S., {Fedynitch}, A., {Palladino}, A., \& {Winter}, W.
  2019, \apjl, 874, L29, \dodoi{10.3847/2041-8213/ab1267}

\bibitem[{{Schlafly} \& {Finkbeiner}(2011)}]{2011ApJ...737..103S}
{Schlafly}, E.~F., \& {Finkbeiner}, D.~P. 2011, \apj, 737, 103,
  \dodoi{10.1088/0004-637X/737/2/103}

\bibitem[{{Science Software Branch at STScI}(2012)}]{2012ascl.soft07011S}
{Science Software Branch at STScI}. 2012, {PyRAF: Python alternative for IRAF}.
\newblock \doeprint{1207.011}

\bibitem[{{Shaw} {et~al.}(2012){Shaw}, {Romani}, {Cotter}, {Healey},
  {Michelson}, {Readhead}, {Richards}, {Max-Moerbeck}, {King}, \&
  {Potter}}]{2012ApJ...748...49S}
{Shaw}, M.~S., {Romani}, R.~W., {Cotter}, G., {et~al.} 2012, \apj, 748, 49,
  \dodoi{10.1088/0004-637X/748/1/49}

\bibitem[{{Stanev} {et~al.}(2000){Stanev}, {Engel}, {M{\"u}cke}, {Protheroe},
  \& {Rachen}}]{2000PhRvD..62i3005S}
{Stanev}, T., {Engel}, R., {M{\"u}cke}, A., {Protheroe}, R.~J., \& {Rachen},
  J.~P. 2000, \prd, 62, 093005, \dodoi{10.1103/PhysRevD.62.093005}

\bibitem[{{Stickel} {et~al.}(1991){Stickel}, {Padovani}, {Urry}, {Fried}, \&
  {Kuehr}}]{1991ApJ...374..431S}
{Stickel}, M., {Padovani}, P., {Urry}, C.~M., {Fried}, J.~W., \& {Kuehr}, H.
  1991, \apj, 374, 431, \dodoi{10.1086/170133}

\bibitem[{{Strotjohann} {et~al.}(2019){Strotjohann}, {Kowalski}, \&
  {Franckowiak}}]{Strotjohann19}
{Strotjohann}, N.~L., {Kowalski}, M., \& {Franckowiak}, A. 2019, \aap, 622, L9,
  \dodoi{10.1051/0004-6361/201834750}

\bibitem[{{Tavecchio} \& {Ghisellini}(2008)}]{2008MNRAS.386..945T}
{Tavecchio}, F., \& {Ghisellini}, G. 2008, \mnras, 386, 945,
  \dodoi{10.1111/j.1365-2966.2008.13072.x}

\bibitem[{{Tody}(1986)}]{1986SPIE..627..733T}
{Tody}, D. 1986, Society of Photo-Optical Instrumentation Engineers (SPIE)
  Conference Series, Vol. 627, {The IRAF Data Reduction and Analysis System},
  ed. D.~L. {Crawford}, 733, \dodoi{10.1117/12.968154}

\bibitem[{{Tody}(1993)}]{1993ASPC...52..173T}
---. 1993, Astronomical Society of the Pacific Conference Series, Vol.~52,
  {IRAF in the Nineties}, ed. R.~J. {Hanisch}, R.~J.~V. {Brissenden}, \&
  J.~{Barnes}, 173

\bibitem[{{van Dokkum}(2001)}]{2001PASP..113.1420V}
{van Dokkum}, P.~G. 2001, \pasp, 113, 1420, \dodoi{10.1086/323894}

\bibitem[{{Vazdekis} {et~al.}(2010){Vazdekis}, {S{\'a}nchez-Bl{\'a}zquez},
  {Falc{\'o}n-Barroso}, {Cenarro}, {Beasley}, {Cardiel}, {Gorgas}, \&
  {Peletier}}]{2010MNRAS.404.1639V}
{Vazdekis}, A., {S{\'a}nchez-Bl{\'a}zquez}, P., {Falc{\'o}n-Barroso}, J.,
  {et~al.} 2010, \mnras, 404, 1639, \dodoi{10.1111/j.1365-2966.2010.16407.x}

\bibitem[{{Zhang} {et~al.}(2020){Zhang}, {Petropoulou}, {Murase}, \&
  {Oikonomou}}]{2020ApJ...889..118Z}
{Zhang}, B.~T., {Petropoulou}, M., {Murase}, K., \& {Oikonomou}, F. 2020, \apj,
  889, 118, \dodoi{10.3847/1538-4357/ab659a}

\end{thebibliography}
\bibliographystyle{aasjournal}

\acknowledgments
Thanks are due to the journal referee for  a constructive criticism.  We thank T. Glauch and F. Oikonomou for fruitful discussions on the interpretation of neutrino alerts. This work was supported by the Initiative and Networking Fund of the Helmholtz Association. We are thankful to \nustar, and \swift~PIs for approving our DDT requests and to the mission operations team for quickly scheduling the observations. Thanks are also due to \nicer~PI for observing the source as a DDT ToO. A.D. acknowledges the support of the Ram{\'o}n y Cajal program from the Spanish MINECO.  We are grateful to staff astronomers Antonio Cabrera at GTC for carrying out OSIRIS observation. This work is based on observations made with the GTC telescope, in the Spanish Observatorio del Roque de los Muchachos of the Instituto de Astrofísica de Canarias, under Director’s Discretionary Time. The work of M.B. is supported through the South African Research Chair Initiative of the National Research Foundation\footnote{Any opinion, finding and conclusion or recommendation expressed in this material is that of the authors and the NRF does not accept any liability in this regard.} and the Department of Science and Innovation of South Africa, under SARChI Chair grant No. 64789. A.G.P. and A.O.G. acknowledge financial support from the Spanish Ministry of Economy and Competitiveness (MINECO) under grant numbers AYA2016-75808-R and RTI2018-096188-B-I00, which are partly funded by the European Regional Development Fund (ERDF). A.O.G. also acknowledges financial support from the Comunidad de Madrid Tec2Space project S2018/NMT-4291. Part of this work is based on results provided by the ASM/RXTE teams at MIT and at the RXTE SOF and GOF at NASA's GSFC.

\vspace{5mm}
\software{XSPEC \citep[][]{Arnaud96}, IRAF \citep[][]{1986SPIE..627..733T,1993ASPC...52..173T}, PyRAF \citep{2012ascl.soft07011S},  HEAsoft (v6.26)}


\end{document}